\documentclass[12pt,nofootinbib,preprintnumbers]{revtex4-1}
\pdfoutput=1
\usepackage{graphicx}
\usepackage{amsmath,amsthm,amssymb,multirow}
\usepackage{xcolor}
\usepackage{caption}
\usepackage{subcaption}
\usepackage[utf8]{inputenc} 
\usepackage{amssymb}
\usepackage{enumerate}
\usepackage{epstopdf}
\usepackage{multirow}
\usepackage{appendix}
\usepackage{url}
\usepackage{caption}
\usepackage{subcaption}
\RequirePackage{snapshot}
\usepackage{wasysym}
\begin{document}

\def\lsim{\mathrel{\rlap{\lower4pt\hbox{\hskip1pt$\sim$}}
    \raise1pt\hbox{$<$}}}
\def\gsim{\mathrel{\rlap{\lower4pt\hbox{\hskip1pt$\sim$}}
    \raise1pt\hbox{$>$}}}
\newcommand{\vev}[1]{ \left\langle {#1} \right\rangle }
\newcommand{\bra}[1]{ \langle {#1} | }
\newcommand{\ket}[1]{ | {#1} \rangle }
\newcommand{\ev}{ {\rm eV} }
\newcommand{\kev}{{\rm keV}}
\newcommand{\mev}{{\rm MeV}}
\newcommand{\gev}{{\rm GeV}}
\newcommand{\tev}{{\rm TeV}}
\newcommand{\pb}{\text{ pb}}
\newcommand{\fb}{\text{ fb}}
\newcommand{\mpl}{$M_{Pl}$}
\newcommand{\mw}{$M_{W}$}
\newcommand{\Ft}{F_{T}}
\newcommand{\Zparity}{\mathbb{Z}_2}
\newcommand{\BLambda}{\boldsymbol{\lambda}}
\newcommand{\be}{\begin{eqnarray}}
\newcommand{\ee}{\end{eqnarray}}
\newcommand{\met}{\;\not\!\!\!{E}_T}
\long\def\/*#1*/{}

\newcommand{\sla}[1]{\setbox0=\hbox{$#1$}           
   \dimen0=\wd0                                     
   \setbox1=\hbox{/} \dimen1=\wd1                   
   \ifdim\dimen0>\dimen1                            
      \rlap{\hbox to \dimen0{\hfil/\hfil}}          
      #1                                            
   \else                                            
      \rlap{\hbox to \dimen1{\hfil$#1$\hfil}}       
      /                                             
   \fi} 

\newcommand{\Tab}[1]{Table~\ref{tab:#1}}
\newcommand{\tab}[1]{table~\ref{tab:#1}}
\newcommand{\tabl}[1]{\label{tab:#1}}
\newcommand{\Fig}[1]{figure~\ref{fig:#1}}
\newcommand{\Figl}[1]{\label{fig:#1}}
\newcommand{\Sec}[1]{section~\ref{sec:#1}}
\newcommand{\Secs}[1]{sections~\ref{sec:#1}}
\newcommand{\Secl}[1]{\label{sec:#1}}
\newcommand{\App}[1]{appendix~\ref{app:#1}}
\newcommand{\Appl}[1]{\label{app:#1}}
\newcommand{\Tabl}[1]{\label{tab:#1}}

\newcommand{\GeV}{\text{ GeV}}
\newcommand{\TeV}{\text{ TeV}}
\newcommand{\draftnote}[1]{{\bf\color{red} #1}}

\title{A New Probe of Dark Sector Dynamics at the LHC}
\author{Arpit Gupta}
\affiliation{Department of Physics and Astronomy, Johns Hopkins
  University, Baltimore, MD 21218, USA}
  
\author{Reinard Primulando}
\affiliation{Department of Physics and Astronomy, Johns Hopkins
  University, Baltimore, MD 21218, USA}
  \affiliation{Department of Physics, Parahyangan Catholic
  University, Bandung, Indonesia}

\author{Prashant Saraswat}
\affiliation{Department of Physics and Astronomy, Johns Hopkins
  University, Baltimore, MD 21218, USA}
\affiliation{Maryland Center for Fundamental Physics, Department of Physics, University of Maryland, College Park, MD 20742}

\preprint{UMD-PP-015-007}
\begin{abstract}
We propose a LHC search for dilepton resonances in association with large missing energy as a generic probe of TeV dark sector models. Such resonances can occur if the dark sector includes a $U(1)$ gauge boson, or $Z'$, which kinetically mixes with the Standard Model $U(1)$. For small mixing, direct $2 \rightarrow 1$ production of the $Z'$ is not visible in standard resonance searches due to the large Drell-Yan background. However, there may be significant production of the $Z'$ boson in processes involving other dark sector particles, resulting in final states with a $Z'$ resonance and missing transverse momentum. Examples of such processes include cascade decays within the dark sector and radiation of the $Z'$ off of final state dark sector particles. Even when the rate to produce a $Z'$ boson in a dark sector process is suppressed, this channel can provide better sensitivity than traditional collider probes of dark sectors such as monojet searches. We find that data from the 8 TeV LHC run can be interpreted to give bounds on such processes; more optimized searches could extend the sensitivity and continue to probe these models in the Run II data.
\end{abstract} 
\maketitle


\section{Introduction}

There is compelling evidence that most of the matter in the Universe is composed of nonbaryonic particles, the dark matter (DM), the nature of which is otherwise unknown. One of the only quantitative data points known about dark matter is its current cosmological energy density, $\Omega_{DM} \approx 0.25$~\cite{Planck:2015xua}. If the dark matter is a relic of freeze-out from the thermal bath of the early Universe, then this relic density can be achieved~\cite{Jungman:1995df} with a dark matter mass of $O(100 \GeV)$ with couplings of order $\sim 0.1$. This ``WIMP miracle'' is a prime motivation for considering dark matter production at the LHC.  Dark matter candidates of weak scale mass can also naturally emerge within theories developed to address the electroweak hierarchy problem, such as supersymmetric models.

Current searches for dark matter at the LHC focus on pair production of invisible DM particles plus radiation from the initial state in the form of jets, photons, or electroweak bosons \cite{Khachatryan:2014rra,Aad:2015zva,Khachatryan:2014rwa,Aad:2013oja,ATLAS:2014wra,Aad:2014tda,Khachatryan:2014tva,Goodman:2010ku,Fox:2011fx,Fox:2011pm,Fox:2012ee,Bai:2012xg,Bell:2012rg,Carpenter:2012rg,Chen:2013gya,Dreiner:2013vla,Askew:2014kqa,Berlin:2014cfa,Petrov:2013nia,Lin:2013sca}. The resulting ``monojet'', ``monophoton'' etc. signatures have considerable SM background, but still allow for constraints to be placed on dark matter interactions with quarks and gluons. Within an effective field theory (EFT) framework, these processes can be correlated with signals from elastic DM scattering in detectors (direct detection) and astrophysical DM annihilation (indirect detection)  \cite{Birkedal:2004xn,Bai:2010hh,Goodman:2010yf,Cheung:2012gi,Arrenberg:2013rzp,Malik:2014ggr}. 


This program is appropriate for the minimal assumption of a single DM particle and no other new physics. However, when one goes beyond this minimal framework, other types of collider searches may provide much more powerful probes \cite{An:2013xka, Bai:2013iqa, DiFranzo:2013vra, Chang:2013oia, Cheung:2013dua, Papucci:2014iwa,Alves:2013tqa,Alves:2015pea,Primulando:2015lfa,Frandsen:2012rk,Chala:2015ama}. A familiar example is that of supersymmetric models, in which searching for squarks and gluinos decaying to a neutralino DM candidate is usually a far more effective probe of the new physics than searches for direct neutralino production. More generally, any new particles associated with dark matter can provide additional collider signatures which may greatly enhance the prospects for discovery. Note that the same enhancement does not extend to direct and indirect detection of dark matter, which are only sensitive to the actual cosmological relics. In this respect colliders provide a unique window into the physics associated with dark matter. 

In this work we specialize to the case where all new particles are gauge singlets under the Standard Model, forming a ``dark sector." Although the model space for such a dark sector is vast, some well-motivated assumptions greatly narrow down the possible collider phenomenology. Consistent with renormalizable field theory, we can consider the new particles to be either fermions, scalars or gauge bosons. Since the coupling of these states to the SM is generally weak, once dark sector particles are produced at colliders they will tend to cascade decay within the dark sector until that is no longer kinematically possible. In particular, particles with arbitrarily weak couplings to the SM can still be produced through decay of or radiation off of other dark sector particles. If the theory preserves baryon and lepton number, or an analogous dark fermion number, then the lightest dark fermion will be absolutely stable and appear as missing energy at colliders. Dark bosons however may not be protected by any quantum numbers and could decay into the Standard Model. Although this decay width may be small due to weak couplings, the branching ratio of the dark boson to the SM can still be large if decays to other states are suppressed-- particularly if they are not kinematically accessible. This allows for visible collider signatures from extremely weakly coupled particles, as in ``hidden valley'' models \cite{Strassler:2006im, Han:2007ae}.  

The collider signatures of a dark gauge boson, or $Z'$, can be particularly striking. The most relevant possible interaction of a dark vector boson with the Standard Model is through the kinetic mixing ``portal'', i.e. the operator $B_{\mu \nu} X^{\mu \nu}$ where $B^{\mu \nu}$ is the SM hypercharge field strength and $X^{\mu \nu}$ is the dark gauge boson field strength \cite{Holdom:1985ag}. As a result of this mixing the $Z'$ will couple to SM currents and thereby decay into pairs of fermions, including an $O(1)$ branching ratio to leptons. This gives a distinctive and easily measured dilepton resonance in collider events. Direct $2 \to 1$ production of $Z'$s has been searched for extensively at colliders \cite{Khachatryan:2014fba, Aad:2014cka, Abazov:2010ti, Aaltonen:2011gp}, but the SM Drell-Yan background limits the sensitivity of such searches to weakly coupled $Z'$s. However, dark sector cascades will tend to produce $Z'$s in association with missing transverse momentum (MET) from invisible dark fermions. In this work we show that leveraging this large MET would allow for searches with very low background and high acceptance for a broad class of dark sector models. We give examples of simplified models, representative of more general possible dark sectors, which could be discovered by such an analysis where current approaches such as monojet searches and inclusive resonance searches fail.     

\begin{figure}
        \centering
        \begin{subfigure}[b]{0.3\textwidth}
                \includegraphics[width=\textwidth]{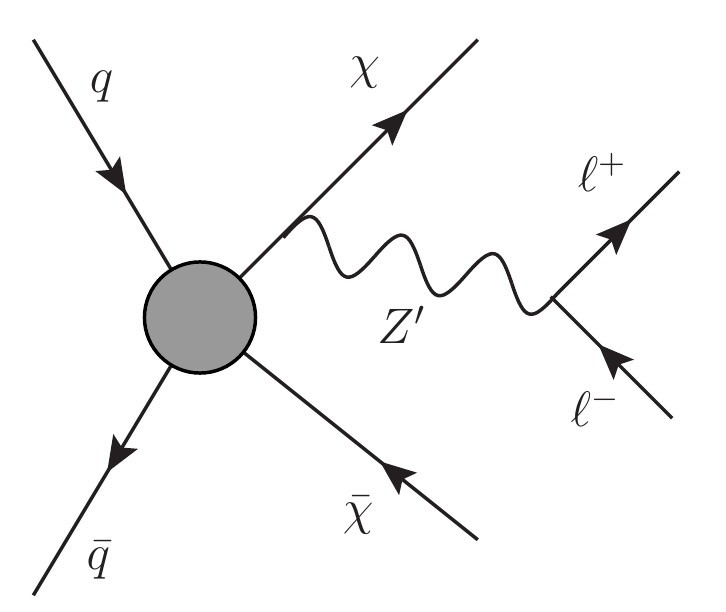}
                \caption{Darkstrahlung}
                \label{fig:darkstrahlungfeynman}
        \end{subfigure}%
        ~ 
        \begin{subfigure}[b]{0.3\textwidth}
                \includegraphics[width=\textwidth]{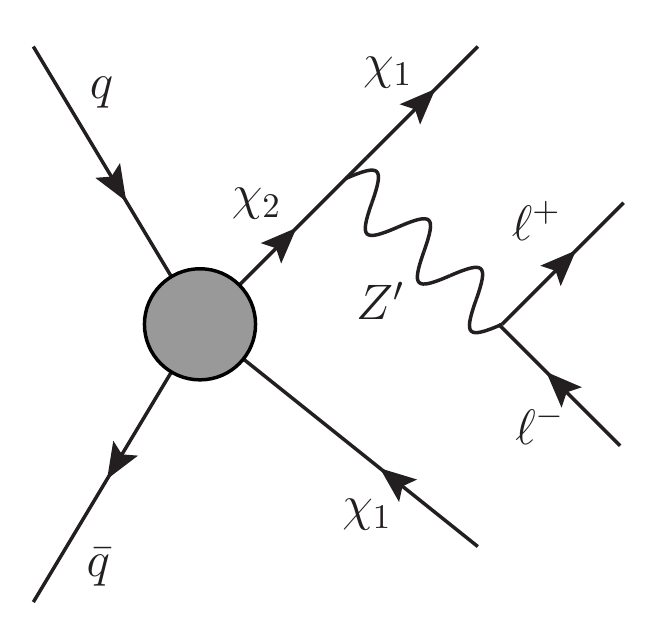}
                \caption{Cascade}
                \label{fig:cascade}
         \end{subfigure}
          ~
         \begin{subfigure}[b]{0.33\textwidth}
                \includegraphics[width=\textwidth]{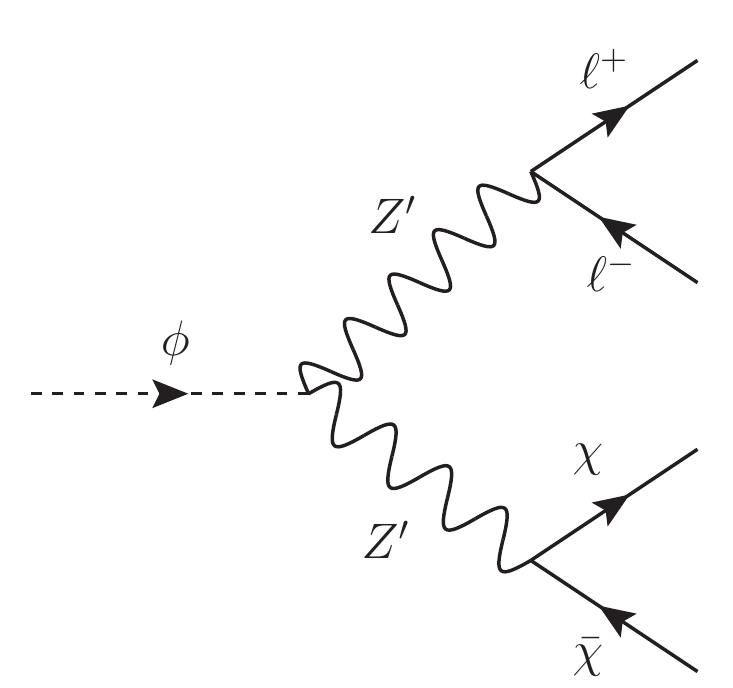}
                \caption{Dark Higgs}
                \label{fig:darkhiggs}
          \end{subfigure}
        \caption{Feynman diagrams for some of the signal processes we discuss in section~\ref{sec:models}. In the process of fig.~\ref{fig:darkstrahlungfeynman}, which we refer to as ``darkstrahlung'' (section~\ref{sec:darkstrahlung}), a $Z'$ boson is radiated off of final state dark sector particles. In~\ref{fig:cascade}, the $Z'$ is produced in the cascade decay of one dark sector particle to another (section~\ref{sec:Cascade}). For both of these processes, we will consider production of the initial dark sector particles through a heavy mediator which can be integrated out. In fig.~\ref{fig:darkhiggs}, an on-shell scalar $\phi$ is produced through mixing with the SM Higgs, and decays to two $Z'$ bosons (section~\ref{sec:H2Zp}).}
         \label{fig:modelfeynman}
\end{figure}

In the next section we describe a proposed search for dilepton resonances plus MET, building off of existing analyses by ATLAS and CMS. In section~\ref{sec:models}, we discuss a variety of signal models (see e.g. figure~\ref{fig:modelfeynman}) which realize the resonance plus MET final state and demonstrate the reach of the proposed search for each of them.

\section{Dileptons plus MET at the LHC}
\label{sec:search}

A search for dilepton resonances in association with missing transverse momentum (MET) can be understood as a straightforward extension of existing, non-resonant LHC studies of dileptons plus MET. Within the SM, analyses in this final states are used to measure the cross section of $W^+W^-$ \cite{Chatrchyan:2013oev, Chatrchyan:2013yaa, ATLAS:2012mec}. Searches in this channel can also probe new physics such as decays of charginos and sleptons in supersymmetric models~\cite{Aad:2014vma}.  In this work, we use cuts building off of the 8 TeV ATLAS search for charginos~\cite{Aad:2014vma}, in particular their signal region $WW$a. The signal region requires two same flavor, opposite charge leptons. The cuts for leptons, missing energy, and jets for the signal region are given in~\Tab{cuts} (all objects are defined after subtraction of tracks and calorimeter energy from pileup interactions, and standard isolation cuts are imposed for leptons). The missing energy variable, ${\met}_{rel}$, is defined as

\begin{equation}
{\met}_{, rel} =\left\{
  \begin{array}{l l}
    \met & \quad \text{if $\Delta \phi_\ell > \pi /2$}\\
    \met \sin\Delta \phi_\ell& \quad \text{if $\Delta \phi_\ell < \pi /2$}
  \end{array} \right.,
\end{equation}
where $\Delta \phi_\ell$ is the azimuthal angle between the $\met$ with the nearest lepton or jets. 

\begin{table}
   \centering
   \parbox{11cm}{
    \begin{ruledtabular}
   \begin{tabular}{c|c} 
   & Cuts \\
   \hline
      $p_{T,e}$ & $> 10$ GeV \\
      $|\eta_e|$ & $< 2.47$\\
      $p_{T,\mu}$ & $> 10$ GeV \\
      $|\eta_\mu|$ & $< 2.4$\\
      $p_{T,\text{leading lepton}}$ & $> 35$ GeV\\
      $p_{T,\text{second lepton}}$ & $> 20$ GeV\\
      $\boldsymbol{m_{\ell\ell'}}$  & $\boldsymbol{> 20}$ \textbf{GeV}, $\boldsymbol{< 120}$ \textbf{GeV}\\
      $\boldsymbol{|m_{\ell\ell'} - m_Z|}$ & $\boldsymbol{> 10}$ \textbf{GeV}\\  
      $\boldsymbol{p_{T,\ell\ell'}}$ & $\boldsymbol{> 80}$ \textbf{GeV}\\  
      $\boldsymbol{{\met}_{rel}}$ & $\boldsymbol{> 80}$ \textbf{GeV}\\  
      jet veto & events with $p^T_j > 20$ GeV and $|\eta_j | < 2.4$\\
      & events with $p^T_j > 30$ GeV and $2.4 < |\eta_j | < 4.5$\\
   \end{tabular}
      \end{ruledtabular}}
   \caption{ The cuts employed by ATLAS for signal region $WWa$ in their search for SUSY electroweakinos in 20~fb$^{-1}$ of 8 TeV data~\cite{Aad:2014vma}. The analysis requires exactly two opposite sign leptons. The signal processes are shown in Fig. \ref{fig:modelfeynman} while the main backgrounds are the leptonic decays of $W^+W^-$ and $t \bar t$. Inside the $Z$ mass window, $ZZ$ process also contributes to the background. In the table above, $m_Z$ is the mass of SM $Z$-boson. ${\met}_{rel}$ is defined in the text. Our proposed searches for $Z'$ + MET signals involve changes to the cuts shown in bold.}
   \Tabl{cuts}
\end{table}

Although the ATLAS signal selection imposes only the cuts on $m_{\ell\ell}$ listed in~\Tab{cuts}, full distributions of the expected and observed $m_{\ell\ell}$ values after all other cuts are shown in ref.~\cite{Aad:2014vma} (see Figure 3a of~\cite{Aad:2014vma} or Fig. \ref{Fig:ATLAS8} of this work). From this information one can derive bounds on resonant dilepton plus MET production for a range of masses. However, the $m_{\ell\ell}$ data as presented is binned in intervals of 10 GeV, which is not optimal for searching for narrow (weakly coupled) resonances. The lepton energy resolution at ATLAS ranges between 1\% to 4\% depending on the lepton flavor and rapidity  \cite{Aad:2011mk, Aad:2014rra}. For CMS, the energy resolution ranges between 1\% to 6\% \cite{Chatrchyan:2013dga, Chatrchyan:2012xi}. Given these resolutions, we expect that the dilepton mass resolution to be better than $\lesssim 3$\% \cite{Aad:2014rra}.
 As a comparison, an ATLAS simulation of Higgs decaying to a dimuon pair \cite{Aad:2014xva} estimates a FWHM of 5.6 GeV ($1\sigma$ resolution of 2.4\%) for $|\eta_\mu| < 1$. For the purpose of this paper, we choose the mass window of $m_{Z'} \pm 2.5\% m_{Z'}$ when estimating the current optimized and future bounds.


Depending on the signal model, altering the cuts on $p_{T,\ell\ell'}$ and ${\met}_{rel}$ could also improve the sensitivity. As we will show in the next section, certain models will tend to produce $Z'$'s with large boosts. For such models, higher values for the $p_{T,\ell\ell'}$ and ${\met}_{rel}$ cuts are more desirable, as one can eliminate backgrounds almost completely without having a significant reduction in signal efficiency.  In other scenarios, however, the $Z'$ can produced with low velocity, so that a looser cut on $p_{T,\ell\ell'}$ and ${\met}_{rel}$ is then necessary to have appreciable signal acceptance. Therefore in a broad search for new resonances in association with missing energy, there is motivation for multiple signal regions with varying cuts. For simplicity, in this work we will always take the same numerical value for the $p_{T,\ell\ell'}$ cut and the ${\met}_{rel}$ cut, and refer to this value as simply the ``MET cut.''

\section{Signal Models}
\label{sec:models}
Before proceeding to specific models of $Z'$ + MET production at the LHC, we review the basic aspects of a kinetically mixed $Z'$. We consider a dark gauge boson of a group $U(1)_D$, which couples to the SM $U(1)_Y$ via the kinetic mixing operator. The Lagrangian of the model is given by 
\begin{equation}
\mathcal L \supset -\frac{1}{4} F'_{\mu\nu} F'^{\mu\nu} + m_{Z'}^2 X_\mu X^\mu + \frac{\epsilon}{2} F'_{\mu\nu} B^{\mu\nu},
\end{equation}
where $ X^\mu$ is the dark gauge boson, $ F'_{\mu\nu} $ is the field strength tensor of $ X^\mu$, and $ B^{\mu\nu}$ is the field strength tensor $U(1)_Y$. These gauge bosons are written in their gauge eigenstates. $m_{Z'}$ is the mass of dark gauge boson, which can arise from various mechanisms, such as spontaneous breaking by a dark Higgs or the Stueckelberg mechanism. The mixing term can be generated by a loop of heavy particles charged under both $U(1)_Y$ and $U(1)_D$. In this scenario, an estimate of the kinetic mixing $\epsilon$ is given by~\cite{Dienes:1996zr}
\begin{equation}
\epsilon \sim \frac{e g_D}{16 \pi^2} \text{log}\left( \frac{m}{\Lambda} \right),
\end{equation} 
where $g_D$ is the $U(1)_D$ coupling constant, $m$ is the mass of the heavy particle that is charged under both gauge groups and $\Lambda$ is a cutoff scale. A wide range of mixing values can be realized in various scenarios: $\epsilon \sim 10^{-12}$ to $\epsilon \sim 10^{-3}$ \cite{Holdom:1985ag, Jaeckel:2010ni, Hewett:2012ns, Essig:2013lka, ArkaniHamed:2008qp, Baumgart:2009tn, Essig:2010ye, Abel:2008ai, Goodsell:2011wn, Goodsell:2009xc}. 

\begin{figure}[tbp]
\begin{center}
\begin{subfigure}[b]{0.45\textwidth}
	\includegraphics[width=\textwidth]{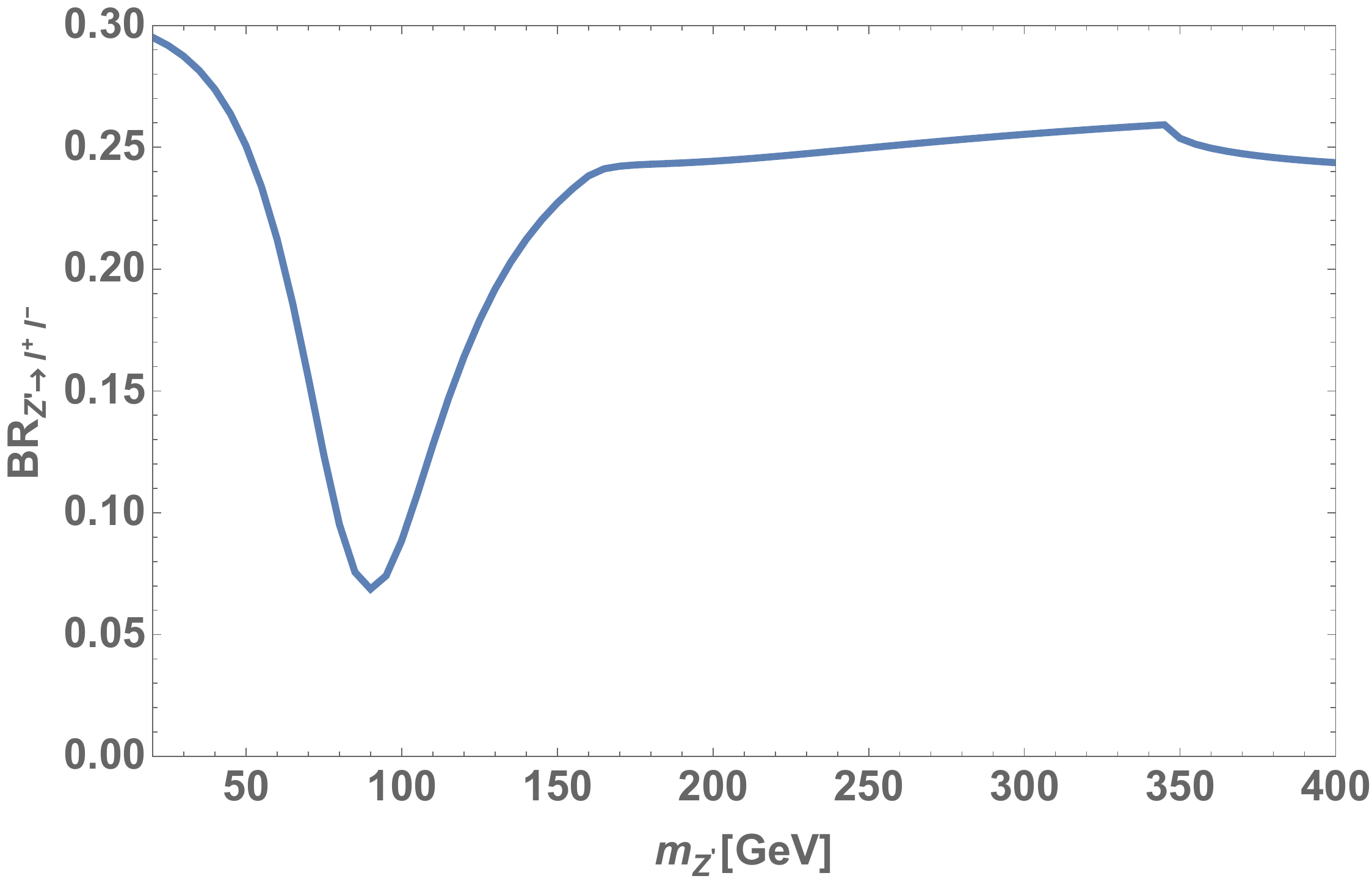}
	\caption{$m_\chi > m_{Z'}/2$}
	\label{Fig:BRv2}
\end{subfigure}
\hspace{5mm}
\begin{subfigure}[b]{0.45\textwidth}
	\includegraphics[width=\textwidth]{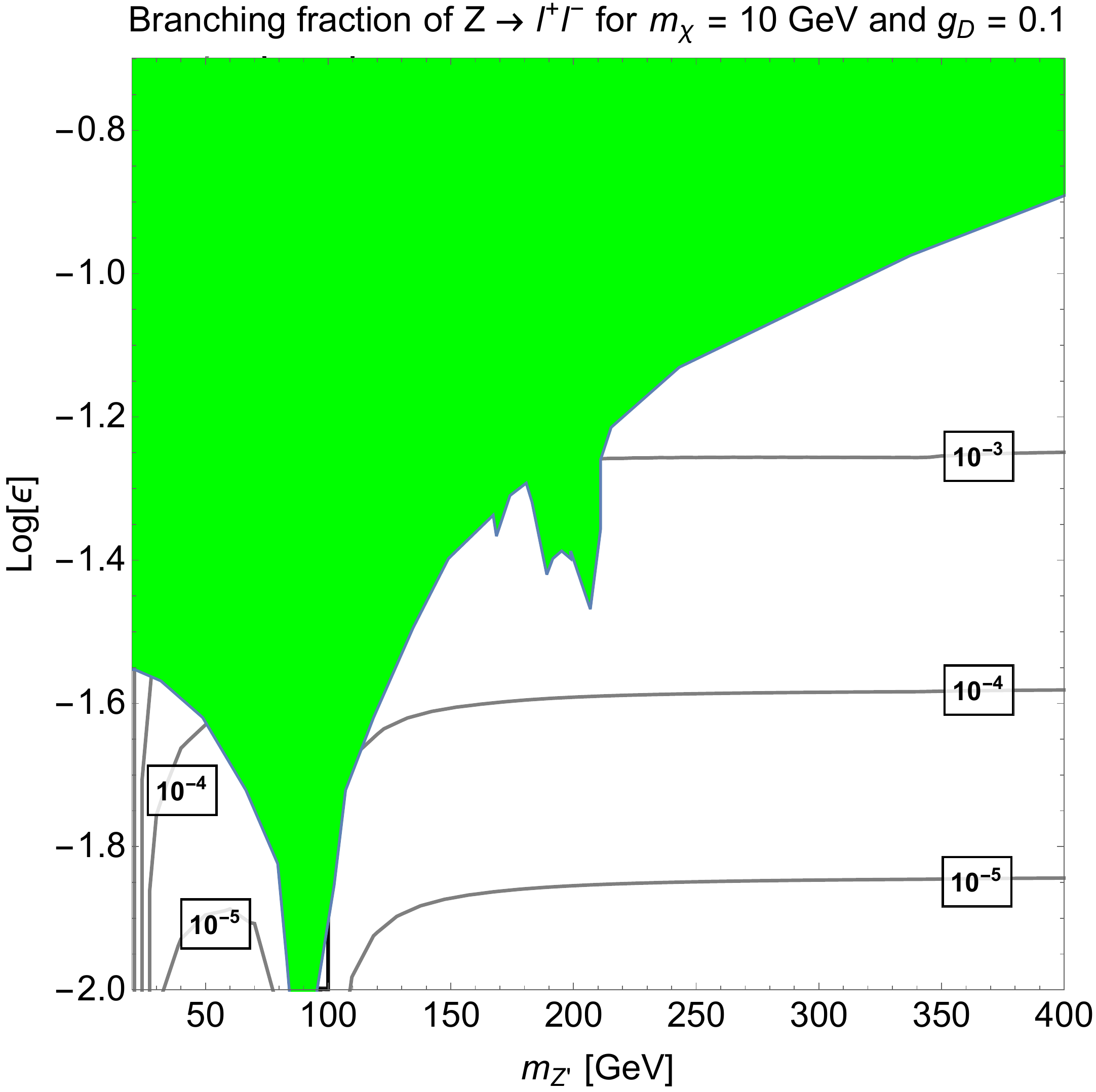}
	\caption{$m_\chi = 10$ GeV}
	\label{Fig:BRepsilonv2}
\end{subfigure}
\end{center}
\caption{Left: Branching fraction of the $Z'$ to two leptons  ($e^+e^-$ and $\mu^+\mu^-$) as a function of mass, assuming the $Z'$ decays only to the SM sector. Right: Contours of the same branching ratio, in the case where there is a 10 GeV dark fermion which the $Z'$ can decay with coupling strength $g_D= 0.1$. The green region on the figure is excluded by precision $Z$ observables at LEP~\cite{Hook:2010tw}.}

\end{figure}

Existing experimental data place various constraints on $Z'$ models. A summary of the bounds for a wide range of masses and kinetic mixings can be found in~\cite{Essig:2013lka, Curtin:2014cca}. However, most of the bounds derived in these references apply only in the case where the $Z'$ decays dominantly to the visible sector. In~\cite{Hook:2010tw}, a model-independent bound is derived by considering the effect of the $Z'$ mixing on precision $Z$ boson measurements at LEP~\cite{2003hep.ex...12023T2,Agashe:2014kda}. For the range of $Z'$ masses that we are concerned with, $m_{Z'} \sim 10 - 1000$ GeV, the model-independent bounds on the mixing parameter are typically $\epsilon \lesssim 0.01 - 0.1$ (green region in figure~\ref{Fig:BRepsilonv2}). 

Given this bound on the mixing, there is still considerable parameter space in which the $Z'$ can decay promptly to the SM. When the $Z'$ decays \textit{only} to the SM sector, the branching fraction to each decay channel is independent of $\epsilon$ and other parameters in the dark sector, and depends only on the $Z'$ mass (for small $\epsilon$). 
In Figure~\ref{Fig:BRv2} we plot the branching fraction of the $Z'$ to leptons ($e^+e^-$ and $\mu^+\mu^-$) in this scenario\footnote{The decay width of $Z'$ to leptons, ignoring the lepton mass, is given by $\Gamma_{Z' \rightarrow \ell^+ \ell^-} =\frac{m_{Z'}}{96 \pi} \left(  \left( g \, c_W  \,s_\alpha - 3 g^\prime \left( s_W s_\alpha - \frac{c_\alpha \epsilon}{\sqrt{1-\epsilon^2}} \right)\right)^{ 2} + \left( g \, c_W  \,s_\alpha + g^\prime \left( s_W s_\alpha - \frac{c_\alpha \epsilon}{\sqrt{1-\epsilon^2}} \right)\right)^{2} \right),$ where $g$ and $g'$ are the SU(2)$_W$ and U(1)$_Y$ coupling constants respectively. $s_W$ and $c_W$ are the sine and cosine of the weak mixing angle respectively. $s_\alpha$ and $c_\alpha$ are the sine and cosine of the mixing angle between the $Z'$, $Z$ and $\gamma$.  In the limit of $\frac{\epsilon s_W}{1-m_{Z'}^2/m_Z^2} \ll 1$, $s_\alpha$ is given by $s_\alpha = \frac{\epsilon s_W}{1-m_{Z'}^2/m_Z^2}$.}.
 
Another possibility is that the $Z'$ can decay to dark sector states. The branching fraction in this case does depend on the kinetic mixing $\epsilon$ and also the details of the dark sector, i.e. the dark gauge coupling constant and the spectrum of states charged under the dark $U(1)$. In Fig. \ref{Fig:BRepsilonv2}, we show contours of the branching fraction of $Z' \rightarrow \ell^+ \ell^-$ in the case of only one light dark sector fermion $\chi$ that the $Z'$ can decay to. As a benchmark point we take $m_\chi = 10$ GeV and $g_D= 0.1$ (and unit charge for the $\chi$).
 
\subsection{Darkstrahlung}
\label{sec:darkstrahlung}

One generic way to produce a $Z'$ in a collider event is to radiate it off of final state dark sector particles (figure~\ref{fig:darkstrahlungfeynman}), a process which we will refer to as ``darkstrahlung.'' To illustrate this, we consider a simplified model of a dark sector fermion $\chi$ that is produced at hadron colliders through an effective operator of the form $\frac{1}{\Lambda^2} \left( \bar{q}  \gamma_\mu \gamma_5 q \right) \left( \bar{\chi} \gamma^\mu \gamma_5 \chi \right)$.\footnote{While the exact Lorentz structure of this operator has little effect on collider phenomenology, if $\chi$ is a cosmological relic then its non-relativistic scattering through this operator generates mostly spin-dependent interactions in direct detection experiments, which are relatively unconstrained.} When initial state radiation is taken into account, such models can be probed by searches for monojets, monophotons, etc.~\cite{Khachatryan:2014rra,Aad:2015zva,Khachatryan:2014rwa,Aad:2013oja,ATLAS:2014wra,Aad:2014tda,Khachatryan:2014tva}. However, if the $\chi$ particles are charged under a $U(1)$ gauge symmetry, then they too can radiate, producing a $Z'$ boson in the final state, with probability determined by the dark $U(1)$ coupling. As we have discussed, such a $Z'$ will always decay to Standard Model states if the dark sector is not kinematically accessible (here, if $m_{Z'} < 2 m_\chi$), leading to an observable final state resonance in association with large missing energy from the stable $\chi$ particles. 

The darkstrahlung process can easily be the dominant production mode of the $Z'$ at colliders. Direct $2 \rightarrow 1$ production of the $Z'$ is suppressed by the kinetic mixing parameter $\epsilon$, and for $\epsilon \lesssim 10^{-2}$ \cite{Jaeckel:2012yz, Hoenig:2014dsa} is too small to be discover amidst the Drell-Yan background. In contrast, the cross-section for the darkstrahlung process depends only on the magnitude of the $\bar{q} q \chi \chi$ operator and the dark $U(1)$ coupling, and the branching ratio for $Z' \rightarrow \ell^+\ell^-$ is independent of $\epsilon$ if dark sector decay modes are inaccessible. Furthermore, the missing energy signature in this channel allows one to almost completely reject Drell-Yan background, greatly increasing the signal sensitivity.

\begin{figure}[tbp]
	\includegraphics[width=.5\textwidth]{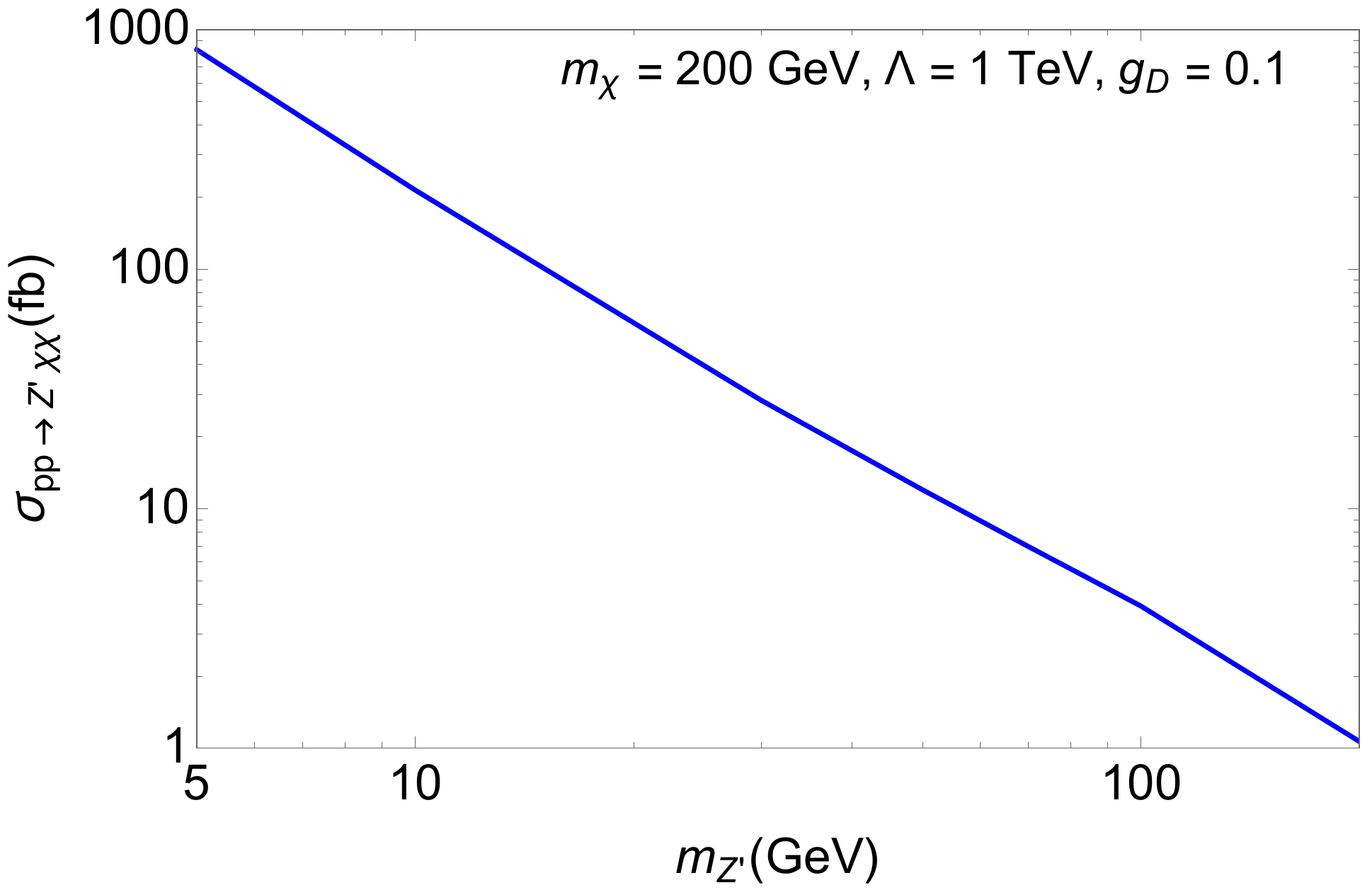}

\caption{The production cross section for $pp \rightarrow \chi\chi Z'$ at the 8 TeV LHC as a function of $m_{Z'}$, for $m_\chi = 200$ GeV, $\Lambda = 1$ TeV and $g_D = 0.1$. (These parameters are just out of reach of current monojet searches, see Figure~\ref{fig:AlphaCompare} and associated discussion.)} 
	\label{Fig:x_sec}
\end{figure}

\begin{figure}[tbp]
	\includegraphics[width=.5\textwidth]{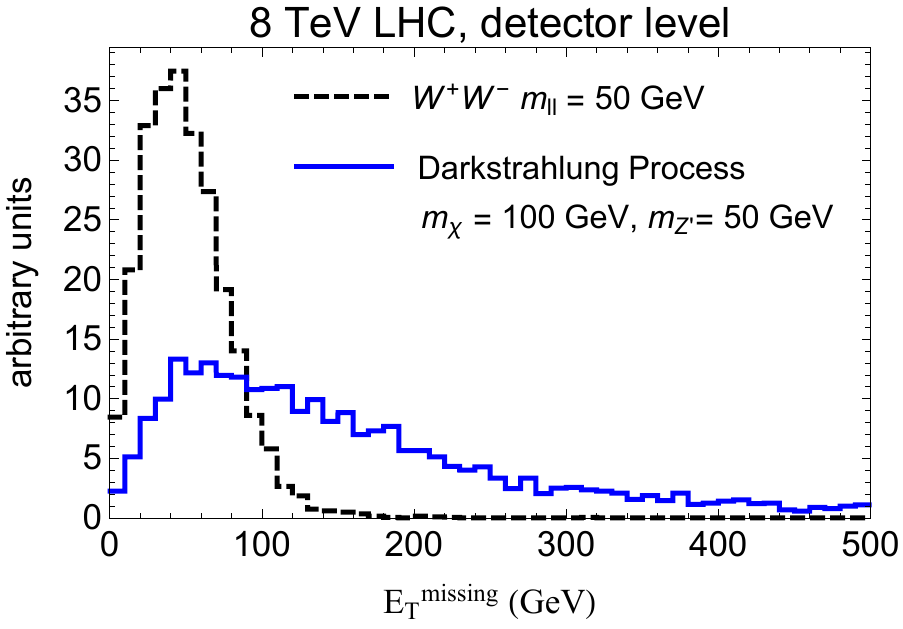}

\caption{Comparison of the MET distributions for the darkstrahlung process with $m_\chi = 200 \GeV$, $m_{Z'} = 50 \GeV$ and the $WW \rightarrow \ell^+\ell^-\nu\bar\nu$ background with the dilepton invariant mass fixed to $m_{\ell\ell'} = 50\pm 2.5 \GeV$. The cuts for both darkstrahlung and $WW$ processes are the same with the ATLAS cuts shown in \Tab{cuts}, except we require $|m_{\ell\ell'} - 50 \GeV| < 2.5 \GeV$ and the ${\met}_{rel}$ cut is not used.} 
	\label{Fig:METcompare2}
\end{figure}

Because the dark matter production proceeds through an operator suppressed by a high scale $\Lambda$, most events occur at high center-of-mass energy $\sqrt{s}$, resulting in the $\chi$ particles having large boosts. The high $\sqrt{s}$ enhances the energy of final state radiation as well, leading to high transverse momenta for the $Z'$.  The rate of FSR includes an enhancement by $s/m_{Z'}^2$ due to the longitudinal mode of the $Z'$ (similar behavior is seen e.g. in bremsstrahlung of electroweak bosons in the annihilation of heavy dark matter particles~\cite{Bell:2008ey,Kachelriess:2009zy}). Figure~\ref{Fig:x_sec} shows the darkstrahlung cross-section as a function of $Z'$ mass for an example set of parameters. Figure~\ref{Fig:METcompare2} shows the MET distribution for darkstrahlung events, compared to Standard Model $WW$ events with the same dilepton invariant mass.

The kinematics of darkstrahlung events depend only on the masses of the $\chi$ and $Z'$ particles; we can therefore interpret collider searches as placing bounds on the darkstrahlung cross-section as a function of these two parameters. As discussed in section~\ref{sec:search}, results from the ATLAS search for electroweakinos can be used to place some constraints on this model. Using the data from Figure 3a of~ \cite{Aad:2014vma}, which is given in dilepton invariant mass bins of 10 GeV, we derive constraints on the darkstrahlung cross-section times branching ratio to $\ell^+\ell^-$ shown in figure~\ref{fig:ATLASconstraints}. (Details of our MC simulation can be found in appendix~\ref{sec:MC}.) A more optimized analysis would use smaller invariant mass bins as well as a slightly higher MET cut to further reduce the SM background. In figure~\ref{fig:OptimizedDarkstrahlung} we show the projected reach for an improved search making use of these tighter cuts, giving an $O(1)$ improvement.

\begin{figure}[tbp]
\makebox[\textwidth][c]{
\begin{subfigure}[b]{0.59\textwidth}
	\includegraphics[width=\textwidth]{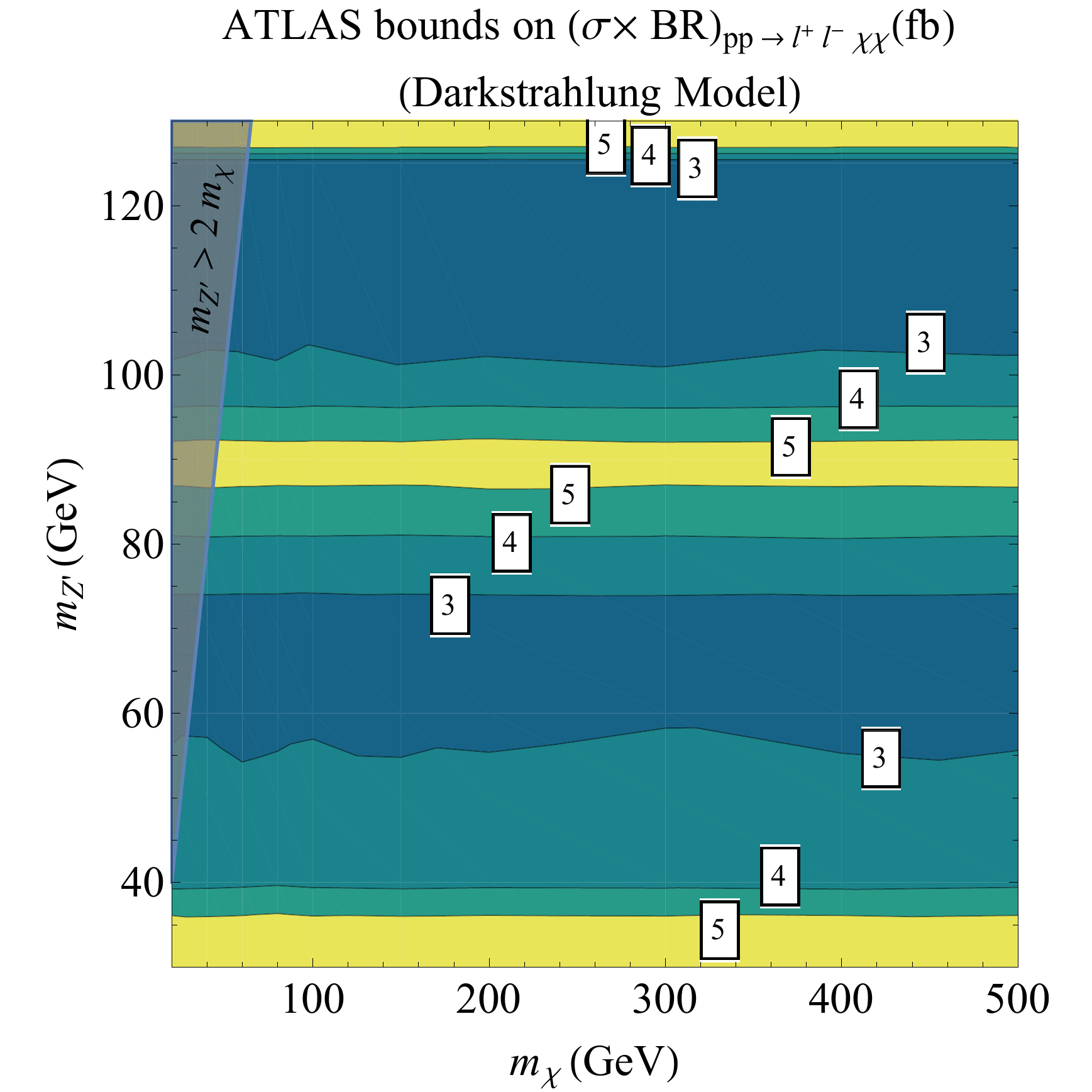}
\caption{}
	\label{fig:ATLASconstraints}
\end{subfigure}
\begin{subfigure}[b]{0.59\textwidth}
	\includegraphics[width=\textwidth]{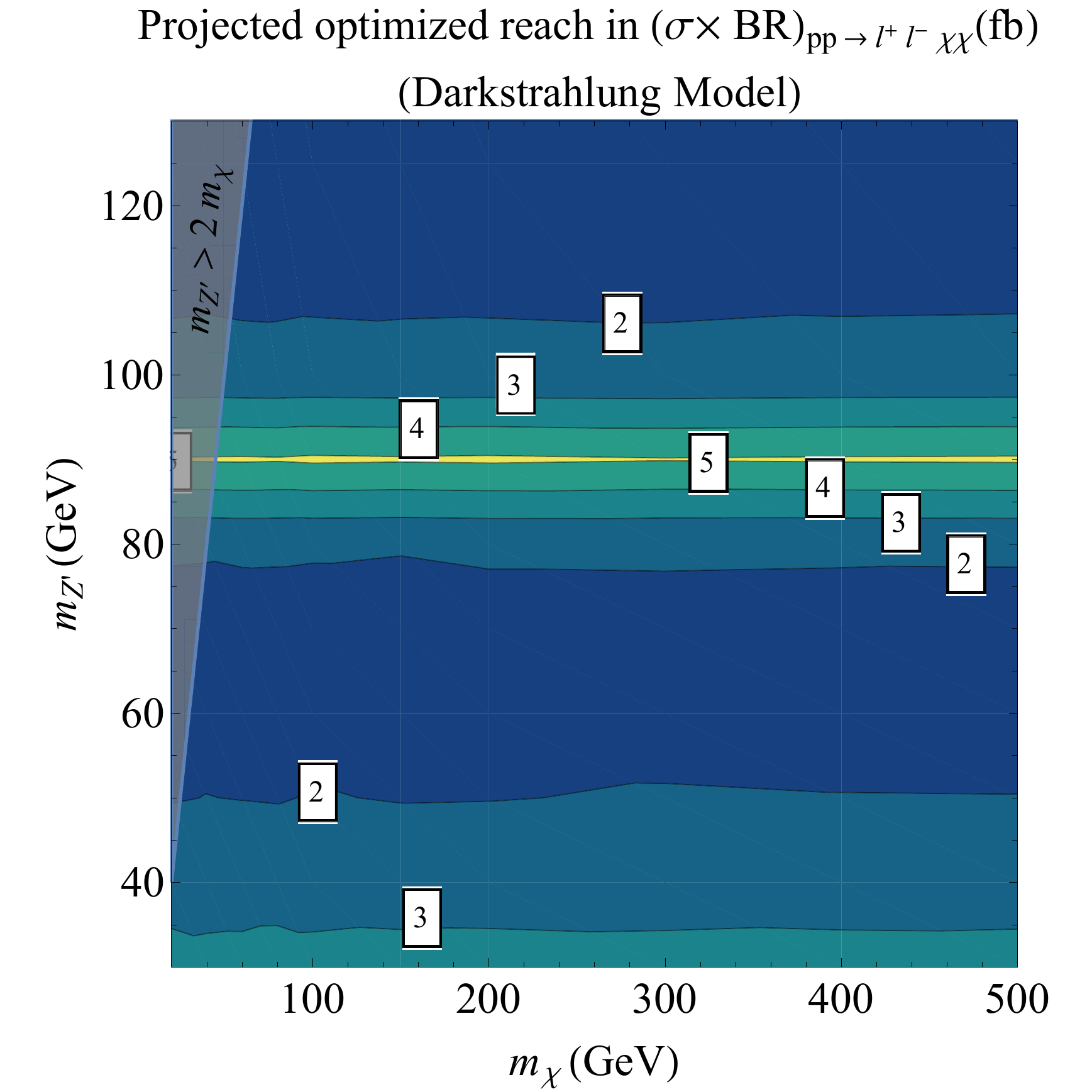}
\caption{}
	\label{fig:OptimizedDarkstrahlung}
\end{subfigure}
}
\vspace{-10mm}
\caption{Left: $95\%$ C.L. constraints on the cross-section times branching ratio of $pp \rightarrow \chi \chi Z'$,  $Z' \rightarrow \ell^+ \ell^-$ for the darkstrahlung model in the plane of $\chi$ mass and $Z'$ mass, from data presented in an ATLAS search for leptons and missing energy (Figure 3a of ref.~\cite{Aad:2014vma}). In the gray shaded region, the $Z'$ is able to decay into $\chi$ particles instead of being forced to decay to SM states. Right: Projected $95\%$ exclusion reach of a more optimized search, with the MET cut increased to 100 GeV and the dilepton invariant mass window narrowed to $\pm 2.5\%$ of the $Z'$ mass.}
\label{fig:Darkstrahlungconstraints}
\end{figure}

\begin{figure}[tbp]
	\includegraphics[width=.7\textwidth]{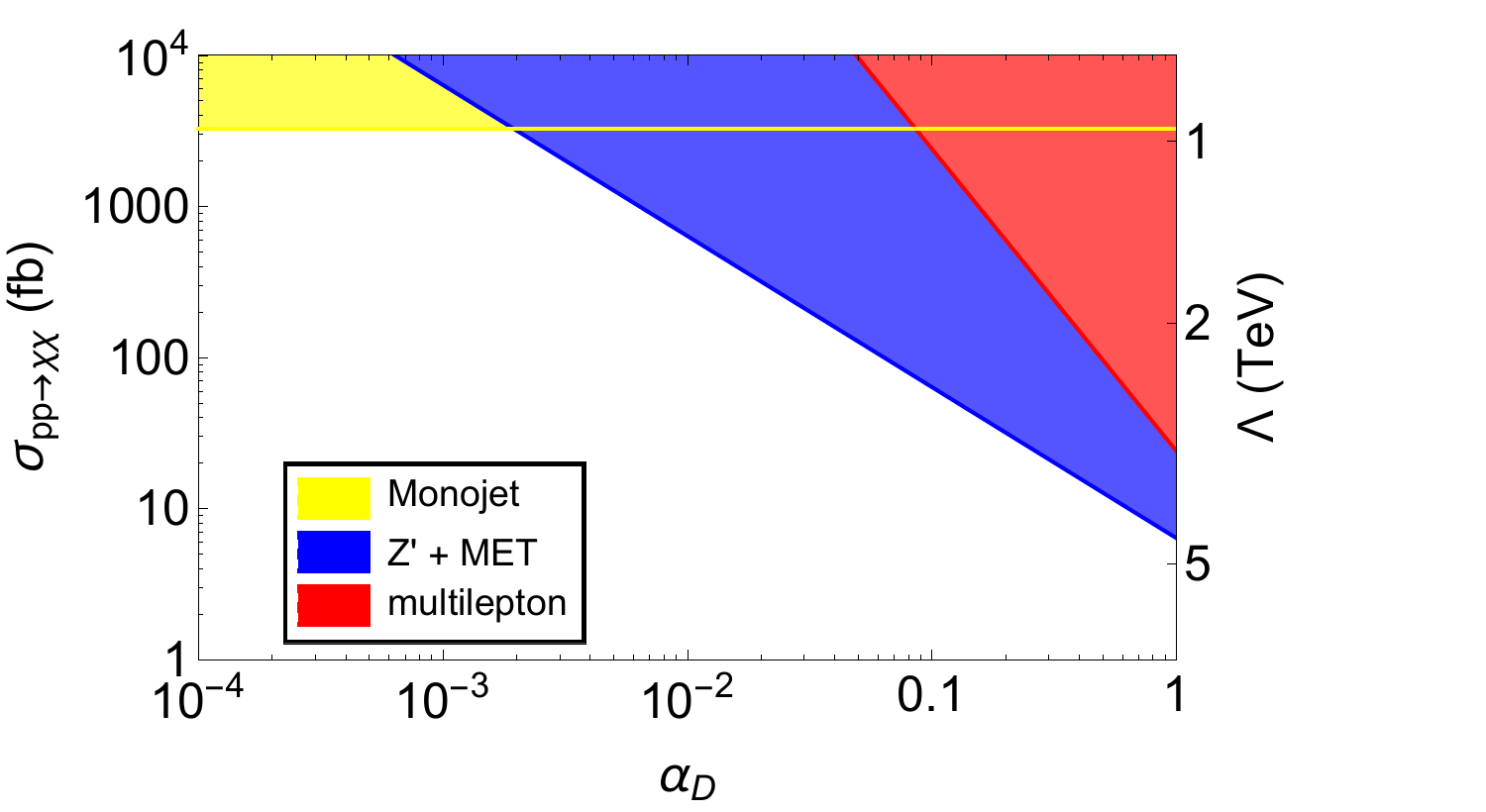}

\caption{Comparison between bounds from monojet \cite{Khachatryan:2014rra}, $Z'$+MET and multi lepton \cite{Chatrchyan:2014aea} as a function of the dark gauge coupling $\alpha_D \equiv g_D^2/(4\pi)$ for  $m_\chi = 200 \GeV$, $m_{Z'} = 120 \GeV$. Multilepton bound is always worse than $Z'$+MET bound for $\alpha_D < 1$.} 
	\label{fig:AlphaCompare}
\end{figure}

In terms of the absolute cross-sections probed, these analyses are far more sensitive to this model than monojet searches, which constrain cross-sections of $O(10 \pb)$. However, the rate for darkstrahlung depends on the dark $U(1)$ coupling and at weak coupling can be much lower than the monojet rate, though even then the darkstrahlung process can be a better probe due to the much greater sensitivity. For large dark $U(1)$ couplings, there can be an appreciable rate to radiate two or more $Z'$'s in a dark sector event, in which case four lepton + MET signatures can be produced. Such final states have close to zero SM background even without reconstructing the $Z'$ resonances~\cite{Chatrchyan:2014aea} and can also be used to place constraints. Figure~\ref{fig:AlphaCompare} shows how the discovery reach of monojet searches, our dilepton + MET proposed search, and multilepton searches compare as a function of the dark $U(1)$ coupling\footnote{The $Z'$+MET bound  shown in Figure~\ref{fig:AlphaCompare} can be approximated as $\sigma_{pp\rightarrow\chi\chi} < (6.8/\alpha_D)$ fb, while the multilepton bound is given by \mbox{$\sigma_{pp\rightarrow\chi\chi} < (23/\alpha_D^2)$ fb.}}. For these parameters, a dilepton + MET search outperforms monojet searches even for $\alpha_D$ as small as $\sim \text{few} \times 10^{-4}$, and is superior to a multilepton search even at strong coupling $\alpha_D \sim 1$.  There is considerable parameter space for which the production operator $\frac{1}{\Lambda^2} \left( \bar{q}  \gamma_\mu \gamma_5 q \right) \left( \bar{\chi} \gamma^\mu \gamma_5 \chi \right)$ can be probed to $\Lambda$ as high as a few TeV. For these values of $\Lambda$, EFT validity at LHC energies is achieved even if the UV theory has order one dimensionless couplings, unlike the parameter space probed by monojet searches~\cite{Busoni:2013lha, Buchmueller:2013dya, Racco:2015dxa}. 

The model we have discussed is presented merely to illustrate possible features of dark sector collider events. A complete model of dark matter would have to address other issues such as the relic abundance. The introduction of a dark $U(1)$ offers a way to produce an appropriate thermal relic abundance; if $m_\chi > m_{Z'}$ then the $\chi$ particles can annihilate to $Z'$'s which then decay to the Standard Model. This can give the correct DM abundance given weak scale masses in the dark sector and couplings of $O(0.1)$ (e.g. $m_\chi = 200 \GeV$, $m_{Z'} = 120 \GeV$, $\alpha_D = 6\times10^{-3}$), realizing the ``WIMP miracle.'' If the $\chi$ is identified as dark matter then it can give direct detection signals mediated by the same EFT operator we used to describe collider production. For Dirac dark matter the vector-channel operator $\left( \bar{q}  \gamma_\mu q \right) \left( \bar{\chi} \gamma^\mu \chi \right)$ is also present and gives large spin-independent direct detection rates, but for Majorana fermions this operator is absent. (In the Majorana case the dark matter mass term can then be generated from a Higgs which breaks the dark $U(1)$.) Of course, all of the results regarding collider physics hold true even if $\chi$ is not the cosmological dark matter, e.g. if it decays to other dark sector particles, if it decays to the SM with a long lifetime, or if it has a small relic abundance. This reflects the fact that collider events can probe the dark sector even beyond dark matter.

\subsection{Cascade decays}
\label{sec:Cascade}

Another potential production mode for $Z'$ bosons is in the decay of one dark sector state to another. A generic dark sector process may involve a number of such decays, with various branching ratios to final states, analogous to the rich set of final states from e.g. top or Higgs production in the Standard Model. As we argued in the introduction, dark sector cascades will typically produce large missing energy from some number of stable invisible particles, but could also result in one or more bosonic resonances, such as a $Z'$, which can decay promptly to the SM. 

Once again, a simple toy model can illustrate the generic phenomenology of this scenario. Consider extending the spectrum of the previous section with an additional dark sector fermion, $\chi_2$, which can decay to a $Z'$ and the lighter fermion, which we denote $\chi_1$ in this section.\footnote{Although the Standard Model famously lacks such ``flavor-changing neutral currents" at tree level, they can occur in more general theories, such as those with vectorlike fermions.} Production of $\chi_1 \chi_2$ or $\chi_2 \chi_2$ can then lead to states with one or more $Z'$'s in association with missing energy, with rates depending on the branching ratio of $\chi_2 \rightarrow Z' \chi_1$. As before, the $Z'$ may be forced to decay to the SM if it is too light to decay to $\chi_1$. Final states with two or more $Z'$ bosons are best probed by existing searches in multilepton channels, as discussed above, so we will focus on the final state with one $Z'$. Final states with one $Z'$ can dominate if the branching ratio of $\chi_2 \rightarrow Z' \chi_1$ is small, or if $\chi_1 \chi_2$ production dominates over $\chi_2 \chi_2$. Both the ``off-diagonal'' production $\chi_1\chi_2$ and the decay $\chi_2 \rightarrow Z' \chi_1$ can naturally arise from spontaneous gauge symmetry breaking in models with Dirac fermions or complex scalars~\cite{Cui:2009xq}.

\begin{figure}[tbp]
	\includegraphics[width=.59\textwidth]{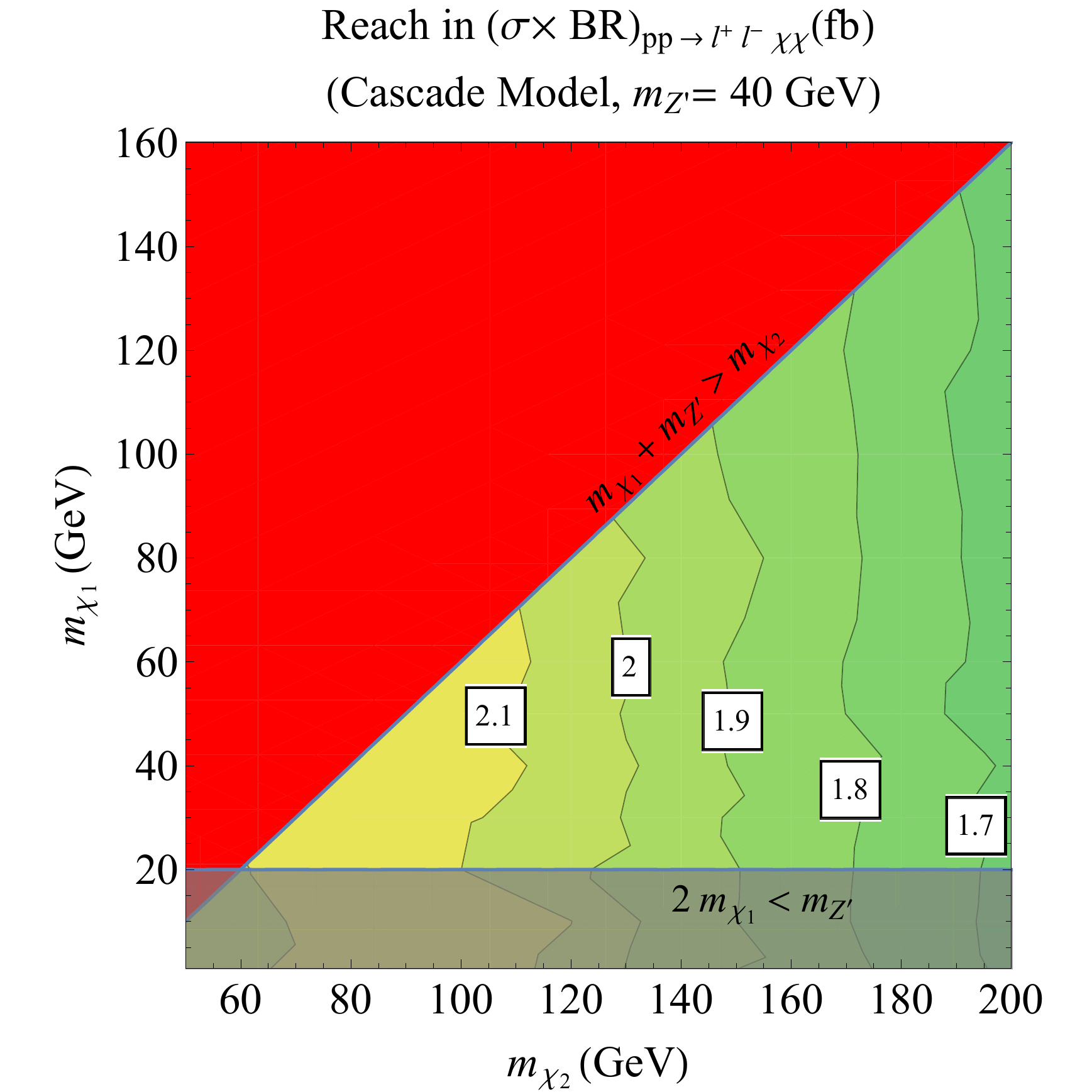}

\caption{Projected $95\%$ exclusion reach in the plane of $\chi_1$ and $\chi_2$ masses for the cascade decay model with $m_{Z'} = 40 \GeV$, from an optimized 8 TeV search as in figure~\ref{fig:Darkstrahlungconstraints}. In the red region $\chi_2 \rightarrow Z' \chi_1$ is kinematically forbidden, while in the gray shaded region the $Z'$ is able to decay into $\chi$ particles instead of being forced to decay to SM states.}
	\label{fig:Cascadeconstraints}
\end{figure}

For the purposes of our Monte Carlo study, we specialize to the case of $\chi_1 \chi_2$ production through the operator $\frac{1}{\Lambda^2} \left( \bar{q}  \gamma_\mu \gamma_5 q \right) \left( \bar{\chi_1} \gamma^\mu \gamma_5 \chi_2 \right)$.  Figure~\ref{fig:Cascadeconstraints} shows the expected reach in cross-section times branching ratio of the optimized 8 TeV search discussed in the previous section, with the $Z'$ mass fixed to 40 GeV while the $\chi_1$ and $\chi_2$ masses are varied. The reach is approximately constant over the entire parameter space shown, even in the ``squeezed limit'' where $m_{\chi_2} \approx m_{\chi_1} + m_{Z'}$, where the $Z'$ is produced nearly at rest in the frame of the $\chi_2$. The invariance of the collider signals with respect to the $\chi$ masses is a result of the large boost of the $\chi$ particles in this production model, which ensures that the $Z'$ from the decay has high $p_T$ independent of the energy released in the decay itself. The cross section bound of $\sim 2 \fb$ corresponds to a value of $\Lambda$ of about $\sim 7-9$ TeV, depending on the mass spectrum.

\subsection{Dark Higgs}
\label{sec:H2Zp}

Thus far we have not discussed the ``dark Higgs'' field $\Phi$ that must be introduced to spontaneously break the dark $U(1)$ gauge symmetry in a weakly coupled model. Such a field can couple to the Standard Model at the renormalizable level through the ``Higgs portal'' operator, $|H|^2|\Phi|^2$. When $H$ and $\Phi$ each acquire their vacuum expectation values, this operator induces mixing between the Standard Model and dark sector Higgses. This allows additional processes to create dark sector states at colliders, with very different kinematics compared to production through the higher-dimension operators discussed above. Nevertheless, we find that searching for resonances accompanied by missing energy can once again probe much of the parameter space of generic models.

\begin{figure}[tbp]
	\includegraphics[width=.5\textwidth]{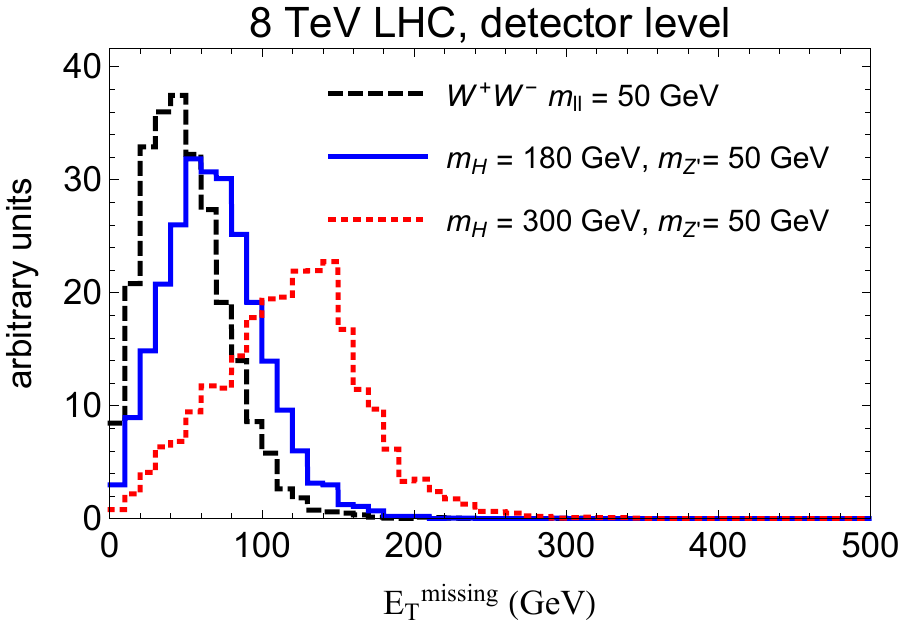}

\caption{Comparison of the MET distributions for the $\phi \rightarrow Z'Z' \rightarrow \ell^+\ell^-\chi\chi$ process with $m_{Z'} = 50 \GeV$ and two different values for the $\phi$ mass, and for the $WW \rightarrow \ell^+\ell^-\nu\bar\nu$ background with the dilepton invariant mass fixed to $m_{\ell\ell'} = 50\pm 2.5 \GeV$. The cuts for both signal and $WW$ processes are the same with the ATLAS cuts shown in \Tab{cuts}, except we require $|m_{\ell\ell'} - 50 \GeV| < 2.5 \GeV$ and the ${\met}_{rel}$ cut is not used.}
	\label{fig:H2ZpMET}
\end{figure}

We will focus on the production of a single on-shell scalar, denoted as $\phi$, which mixes with the SM Higgs. This state can then be produced through all of the same channels as the SM Higgs; for simplicity we will focus on the dominant gluon-fusion-initiated $2 \rightarrow 1$ process.  Once produced, $\phi$ may be able to decay to two $Z'$ bosons, depending on the parameters of the model. If these $Z'$ states are kinematically forced to decay to the Standard Model (as we assumed in the previous subsection), then there can be spectacular four-lepton final states reconstructing multiple resonances, as considered in ~\cite{Curtin:2013fra, Gopalakrishna:2008dv, Lee:2013fda, Martin:2011pd, Chang:2013lfa}. However, if there are dark sector particles which the $Z'$ can decay into, then the branching ratio to SM states becomes $\sim \frac{\epsilon^2 g_Y^2}{g_D^2}$, which can be quite low. The rate for four-lepton final states may then be negligibly small. However, there is a much larger rate for only one $Z'$
 decay to the SM while the other decays to the dark sector, giving a resonance plus MET signal.   

\begin{figure}[tbp]
\makebox[\textwidth][c]{
\begin{subfigure}[b]{0.59\textwidth}
	\includegraphics[width=\textwidth]{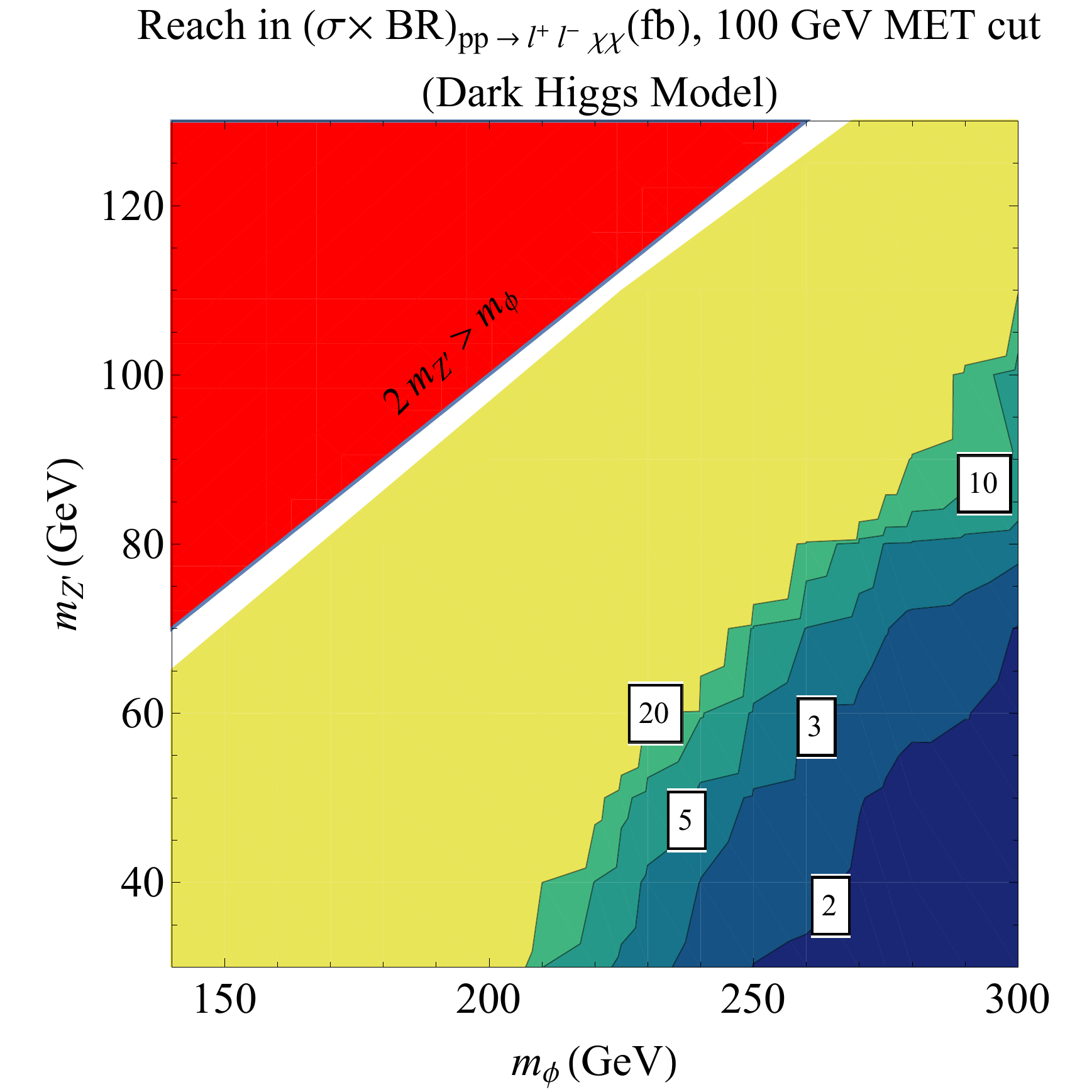}
\caption{}	
\label{fig:H2ZpMET100}
\end{subfigure}
\begin{subfigure}[b]{0.59\textwidth}
	\includegraphics[width=\textwidth]{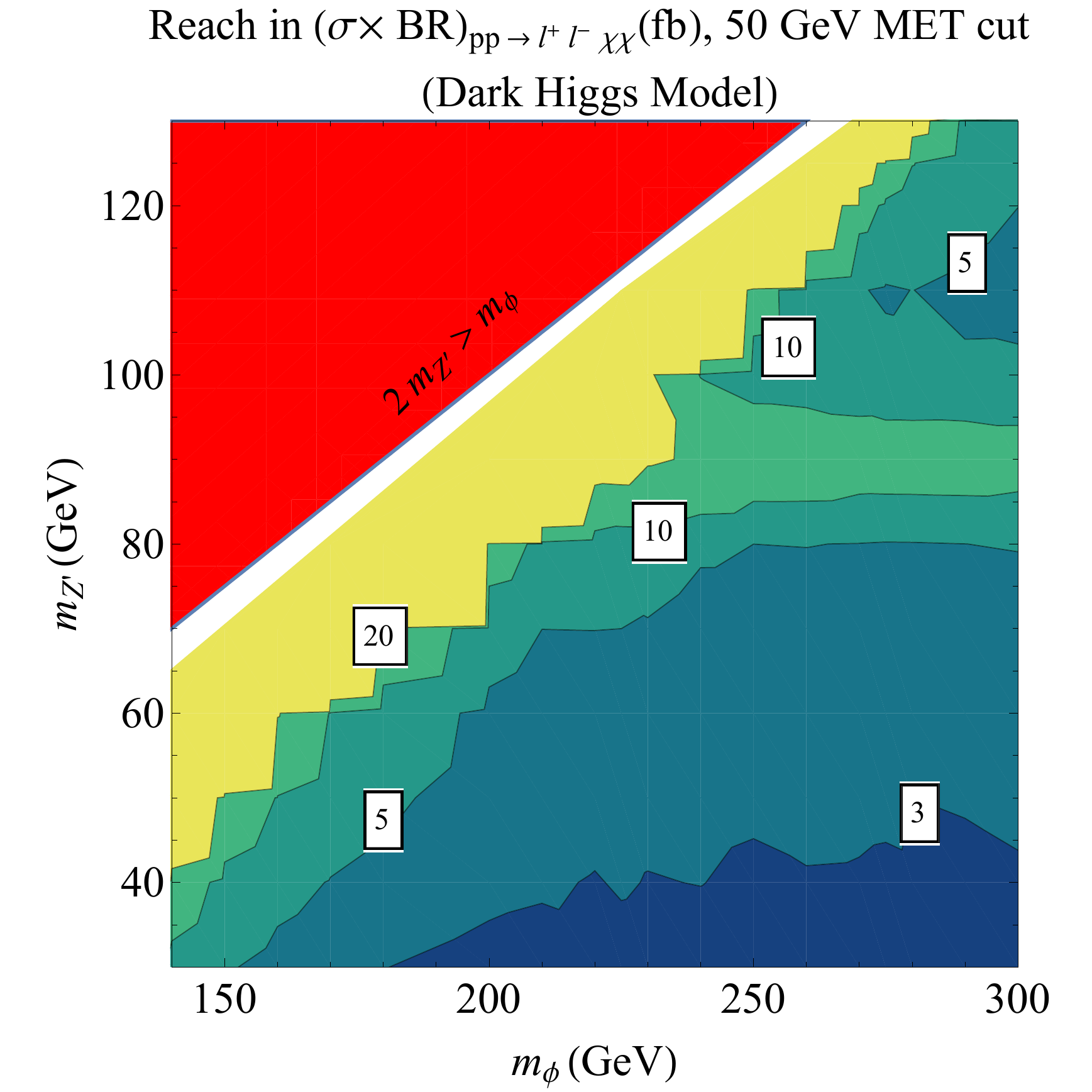}
\caption{}		
\label{fig:H2ZpMET50}
\end{subfigure}
}
\vspace{-10mm}
\caption{Left: Projected $95\%$ exclusion reach in the plane of the dark Higgs and $Z'$ masses for the dark Higgs model, from an 8 TeV search as in figure~\ref{fig:Darkstrahlungconstraints}, with a 100 GeV MET cut. In the red region, the decay $\phi \rightarrow Z'Z'$ is kinematically forbidden. Right: Projected reach for the same search with the MET cut reduced to 50 GeV.} 
\label{fig:H2Zpconstraints}
\end{figure}

The MET distributions for this signal model are compared to the $WW$ background in figure~\ref{fig:H2ZpMET}, for two different choices of parameters. The differing cut-offs for the signal distributions motivate multiple signal regions with different MET cuts to maximize sensitivity over all of parameter space. As above, using MC simulation we can estimate the reach of our proposed search in terms of the total cross-section times branching ratio for the $\phi \rightarrow Z'Z' \rightarrow \ell^+\ell^-\chi\chi$ process, as a function of the $\phi$ and $Z'$ masses. We show the results in figure~\ref{fig:H2Zpconstraints} for two different MET cuts, a 100 GeV cut as considered thus far (Fig.~\ref{fig:H2ZpMET100}) and 50 GeV cut (Fig.~\ref{fig:H2ZpMET50}). 

\begin{figure}[tbp]
\makebox[\textwidth][c]{
\begin{subfigure}[b]{0.59\textwidth}
	\includegraphics[width=\textwidth]{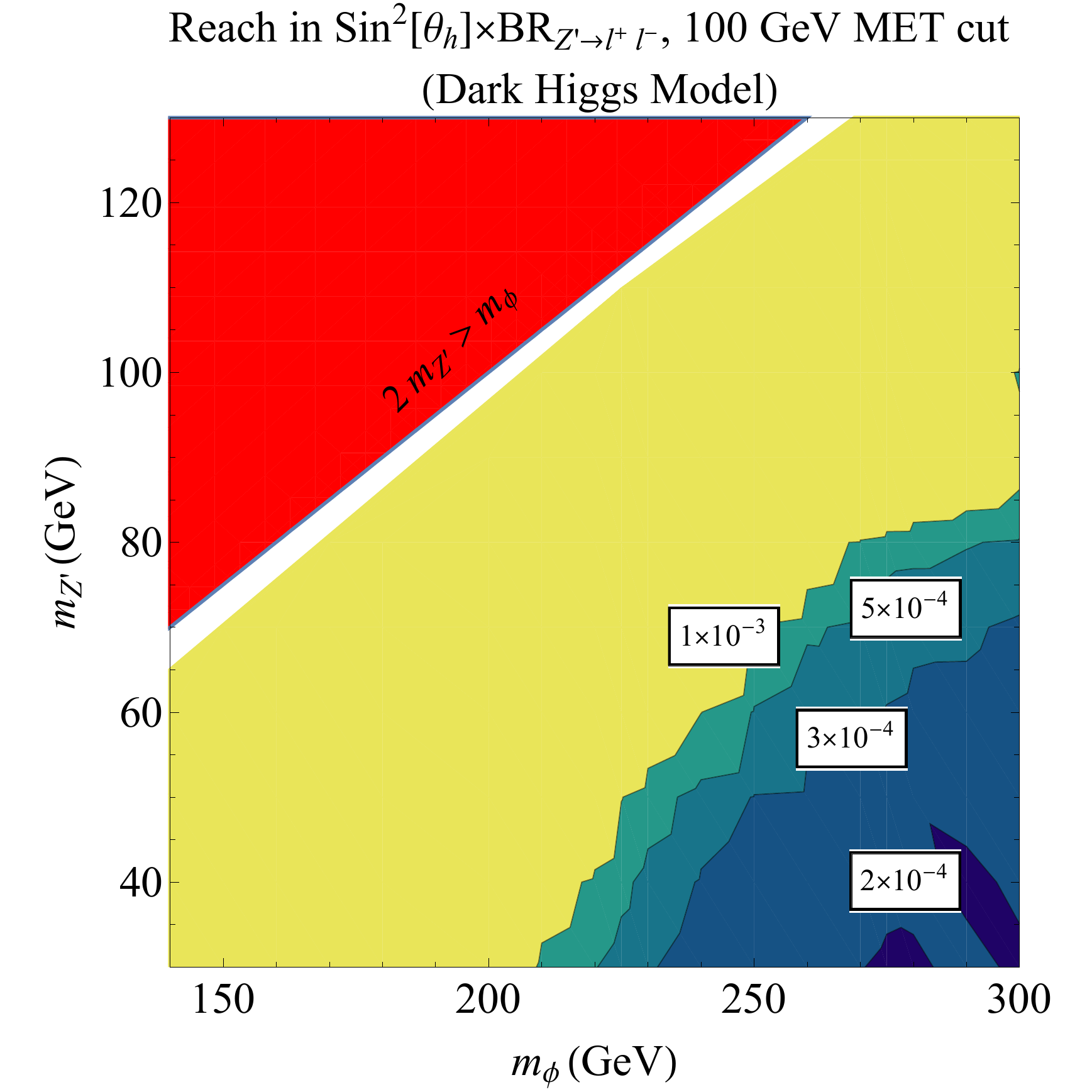}
\caption{}	
\label{fig:H2ZpBRMET100}
\end{subfigure}
\begin{subfigure}[b]{0.59\textwidth}
	\includegraphics[width=\textwidth]{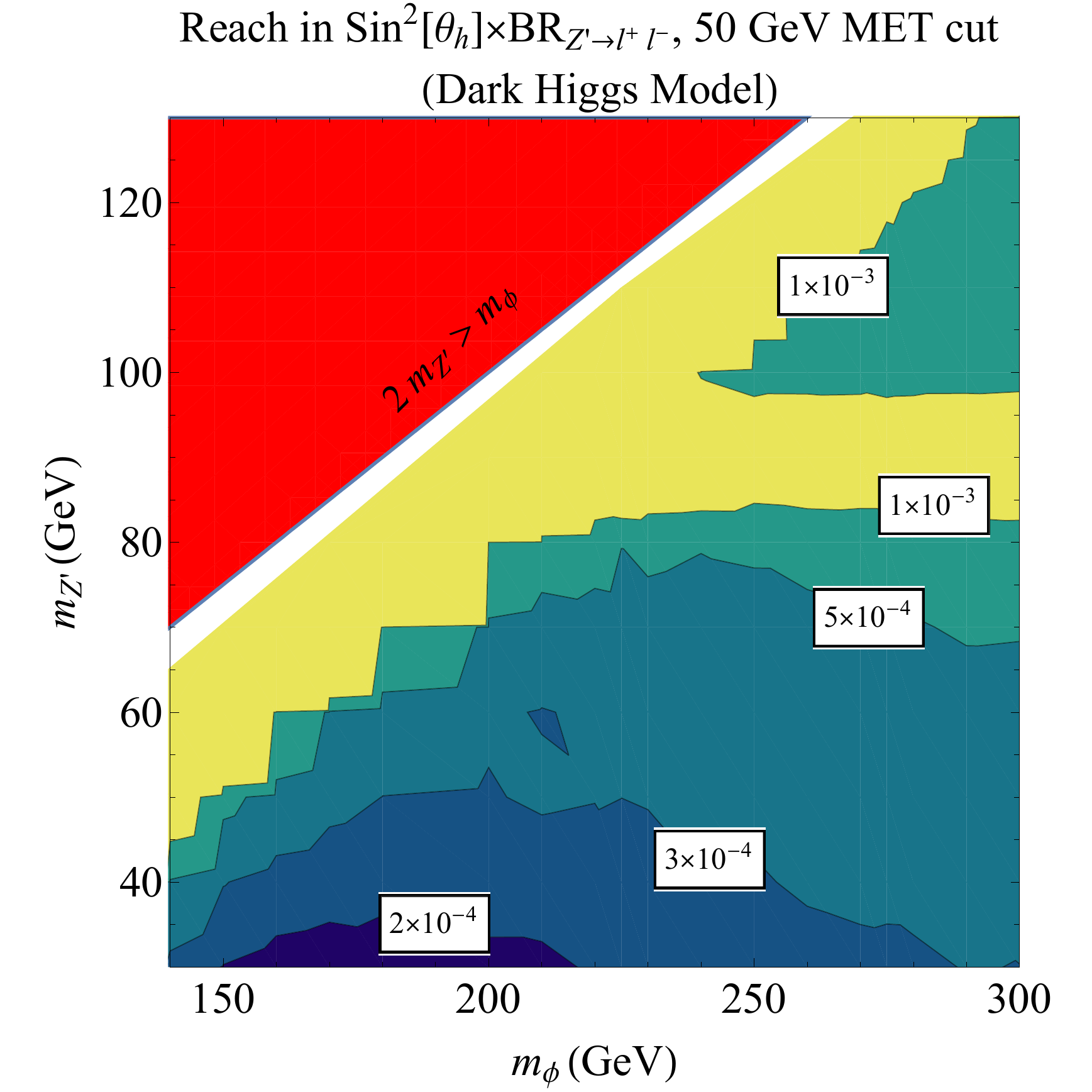}
\caption{}	
\label{fig:H2ZpBRMET50}
\end{subfigure}
}
\vspace{-10mm}
\caption{Left: Projected $95\%$ exclusion reach on the quantity $\sin^2 \theta_h \times \text{BR}(Z' \rightarrow \ell^+ \ell^-)$ for the dark Higgs model, from an 8 TeV search as in figure~\ref{fig:Darkstrahlungconstraints}, with a 100 GeV MET cut. Right: Projected reach for the same search with the MET cut reduced to 50 GeV.} 
\label{fig:H2zpBRconstraints}
\end{figure}

In this model the production cross-section of the dark Higgs is determined completely by the mixing angle $\theta_h$ between the dark and SM Higgs states. We can therefore convert the above bounds on total cross-section times branching ratio for $pp \rightarrow \ell^+\ell^- \chi \chi$ to bounds on $\sin^2 \theta_h \times \text{BR}(Z' \rightarrow \ell^+ \ell^-)$ (with the approximation $\text{BR}(\phi \rightarrow \ell^+ \ell^- \chi \chi) \approx 2 \times \text{BR}(Z' \rightarrow \ell^+ \ell^-)$). The resulting bounds are shown in figure~\ref{fig:H2zpBRconstraints}, again for two different choices of MET cut. Values of order few times $10^{-4}$ as would be probed by this search could be achieved e.g. with $\sin^2 \theta_h \approx 0.1$ (as allowed by Higgs coupling fits~\cite{Khachatryan:2014jba, Aad:2013wqa}) and $\epsilon \approx 0.1 g_D$.

\subsection{Twin Higgs}
\label{sec:TwinHiggs}

Beyond providing dark matter candidates, specific dark sector models can alleviate the hierarchy problem of the SM, if they include states which couple to the Higgs so as to cancel the SM Higgs quadratic divergences. This paradigm of ``neutral naturalness,'' which includes models such as Twin Higgs~\cite{Chacko:2005pe} (more generally ``Orbifold Higgs"~\cite{Craig:2014aea,Craig:2014roa}) and folded supersymmetry~\cite{Burdman:2006tz}, can explain the smallness of the electroweak scale without introducing new light fields with SM color charge, which thus far have not been observed. In the Twin Higgs model~\cite{Chacko:2005pe}, for example, the SM Higgs fields are Goldstone bosons of a global $SU(4)$, which is explicitly broken to a gauged $SU(2)_A \times SU(2)_B$, where the $SU(2)_A$ sector is identified with the Standard Model $SU(2)$ sector. If the $SU(2)_B$ gauge sector is an exact $Z_2$ copy of the $SU(2)_A$ sector, then quantum corrections to the quadratic terms in the scalar potential respect the full $SU(4)$ global symmetry and therefore give no contributions to the potential of the SM Higgs. Appropriately controlled breaking of the $Z_2$ symmetry can give realistic phenomenology without reintroducing Higgs mass divergences at a problematic level. 

Twin Higgs theories therefore require a dark sector with a multitude of states. The exact spectrum of this sector can vary depending on the sources of $Z_2$ symmetry breaking.  The dark sector can be very difficult to probe at colliders, as it couples to the SM mainly through the Higgs boson and mostly gives rise to invisible final states. However, our generic argument that dark sector models offer the possibility of a dilepton resonance + MET signature applies in this scenario as well, offering a potential new probe. A kinetic mixing term between the $U(1)$ fields of the two sectors respects the $Z_2$ (and could easily be generated at high scales), and depending on the spectrum of the model could give a significant rate for dark sector gauge bosons to decay to the SM. 

As a specific example, we will work within the original Twin Higgs model~\cite{Chacko:2005pe}. To discuss the phenomenology however we must specify the sources of $Z_2$ breaking. Since the top Yukawa coupling and the $SU(2) \times U(1)$ gauge couplings give the largest contributions to the Higgs quadratic divergence, the corresponding couplings in the dark sector must respect the $Z_2$ symmetry almost exactly. However, we can break the $Z_2$ for fields that are weakly coupled to the Higgs without upsetting naturalness (as in e.g the ``Fraternal Twin Higgs''~\cite{Craig:2015pha}). This is necessary to achieve realistic phenomenology, as having many light fields in the dark sector such as the twin electrons, neutrinos etc. is in tension with constraints from Big Bang Nucleosynthesis (BBN) as well as cosmic microwave background (CMB) observations, if the two sectors reach thermal equilibrium (see e.g.~\cite{Barbieri:2005ri}). One way this can be alleviated is if there is only one generation of fermions in the dark sector, corresponding to the third SM generation. (A complete generation ensures anomaly cancellation.) Furthermore, the Yukawa couplings of the lighter fermions, the ``dark bottom'' and ``dark tau'', can be larger than in the SM without contributing too much to the Higgs mass. A mass for the dark sector photon also helps reduce tension with BBN/CMB bounds. This can be accomplished if we introduce an explicit mass for the dark sector $U(1)$ (hypercharge) field, denoted $B'$; this soft breaking will not affect cancellation of quadratic divergences. In the mass basis, we will denote the heavier and lighter neutral dark gauge bosons as $Z'$ and $A'$ respectively; in the limit $m_{B'} = 0$ these become the exact $Z_2$ partners of the SM $Z$ boson and photon.

The most accessible dark sector state at colliders is the dark Higgs, which mixes with the SM Higgs as in the model of the previous section. Unlike in the SM, the dark Higgs has a ``mixed'' decay to a $Z'$ and $A'$ when $m_{B'} \neq 0$, as the mass basis and Higgs coupling basis no longer coincide. (For small $m_{B'}$, the rate for decay to two $A'$'s is much lower.) The subsequent decays of the $Z'$ and $A'$ then depend on the fermionic spectrum of the dark sector. One possibility is that the lighter $A'$ boson is kinematically constrained to only decay to the SM through kinetic mixing, while the heavier $Z'$ can decay into dark fermions (generically denoted by $\chi$). In this scenario dark Higgs production leads to the distinctive resonance + MET final state, $\phi \rightarrow Z'A'$, $A' \rightarrow \ell^+\ell^-$, $Z' \rightarrow \chi\chi$.

The relevant physical parameters to describe the production and decay rates for this process are the dark Higgs mass, the mixing angle $\theta_h$ between the two Higgs fields, the $Z'$ mass, and the $A'$ mass. These four parameters plus the SM Higgs mass and vev are determined by the Lagrangian parameter $m_{B'}$ and the five parameters of the (renormalizable) SM plus dark sector Higgs potential.~\footnote{For a Twin Higgs mode to be completely natural, there must be an explanation for the smallness of the scale $f$ where the global $SU(4)$ (and gauged $SU(2)_B$) is broken, in terms of e.g. strong dynamics or supersymmetry (as in~\cite{Falkowski:2006qq, Chang:2006ra, Craig:2013fga}) near the scale $f$. Such a UV completion may constrain the parameters of the potential or introduce additional non-renormalizable terms. Not being committed to a particular UV model, we ignore such effects.} The dark Higgs branching ratio to $Z' A'$ can also be computed from these inputs, so the full rate for $\phi \rightarrow Z'A'$, $A' \rightarrow \ell^+\ell^-$, $Z' \rightarrow \chi\chi$ is predicted given the physical masses and the Higgs mixing. (We approximate the $Z'$ branching ratio to dark fermions as unity.) Typical branching ratios for the $\phi$ to this final state are of order $10^{-3}$.

\begin{figure}[tbp]
\makebox[\textwidth][c]{
\begin{subfigure}[b]{0.59\textwidth}
	\includegraphics[width=\textwidth]{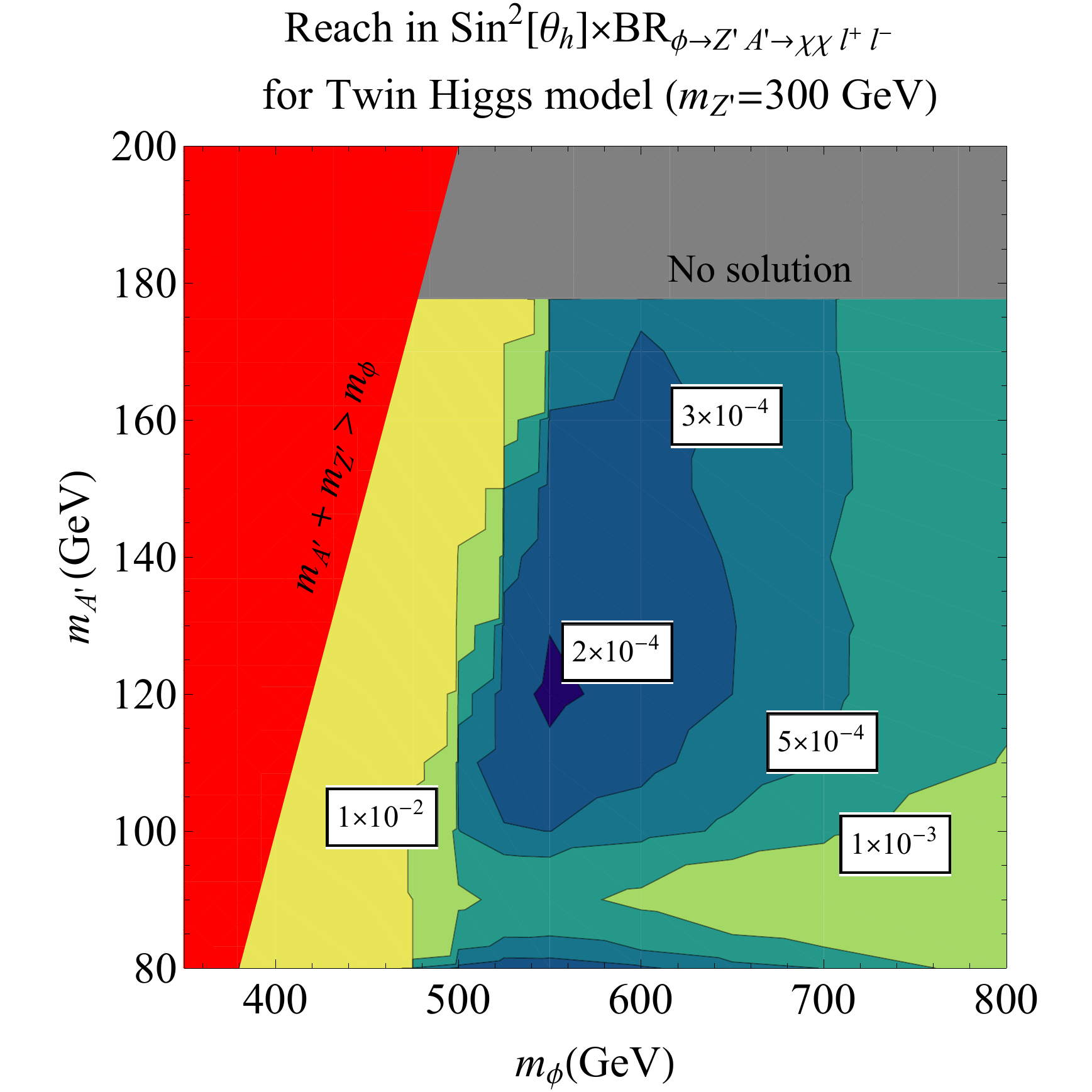}
\caption{}	
\label{fig:TwinHiggsConstraintsBR300}
\end{subfigure}
\begin{subfigure}[b]{0.59\textwidth}
	\includegraphics[width=\textwidth]{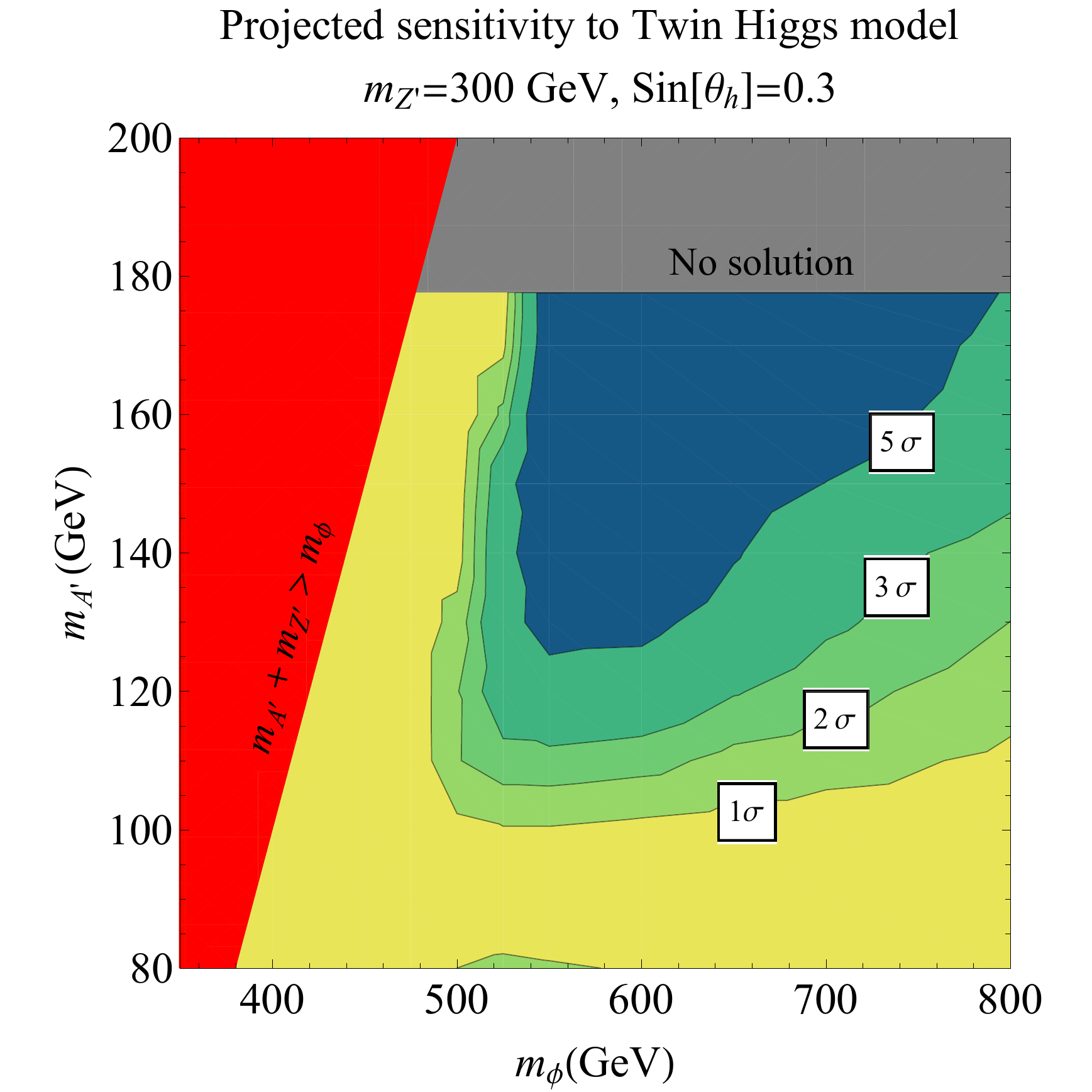}
\caption{}	
\label{fig:TwinHiggsConstraints300}
\end{subfigure}
}
\vspace{-10mm}
\caption{Left: Projected $95\%$ exclusion reach on $\sin^2 \theta_h \times \text{BR}(\phi \rightarrow Z' A' \rightarrow \chi \chi \ell^+ \ell^-)$ for the minimal Twin Higgs model, projected for a search in $100 \fb^{-1}$ of 14 TeV data using a MET cut of 120 GeV.  The $Z'$ mass is fixed to 300 GeV. Right: Expected discovery significance for the process $\phi \rightarrow Z'A', A' \rightarrow \ell^+\ell^-, Z' \rightarrow \chi\chi$ in the same search, with $m_{Z'}$ set to 300 GeV and $\sin \theta_h$ set to 0.3. At each point in the $(m_\phi, m_{A'})$ space, the parameters of the model are uniquely determined, so the branching ratios can be computed.}
\label{fig:TwinHiggsConstraints}
\end{figure}


The large masses required in this model and the small branching ratio to the desired final state limit the discovery potential at the 8 TeV LHC.  With sufficient luminosity however, 14 TeV LHC data can begin to probe this process. As in the previous sections, we simulate the signal and the $WW$, $t\bar{t}$ and $ZV$ backgrounds at 14 TeV in the presence of pileup, now with 40 pileup events expected per crossing~\cite{CMS:2013xfa5}. We impose the same cuts as in the 8 TeV search, except for an increased MET cut of 120 GeV and a modified pileup subtraction algorithm (see Appendix~\ref{sec:MC}). Figure~\ref{fig:TwinHiggsConstraints} shows the discovery reach in $100 \fb^{-1}$ of 14 TeV data, in the plane of the dark Higgs and $A'$ mass, for $m_{Z'} = 300 \GeV$. Sizable regions of parameter space can be probed at the LHC through this technique, with possible improvements from refinements of the search strategy.

Our analysis here is based on what is essentially still a toy model, as the minimal Twin Higgs requires UV completion in order to be truly natural (or for its fine-tuning to even be defined). However, our results do suggest that it may be possible to discover the dark sector states of Twin Higgs theories by searching for dilepton resonances in association with missing energy. Indeed, if Nature stabilizes the electroweak scale through a dark sector, then processes of this type are one of very few ways to directly probe the spectrum of the theory.

\section{Conclusion}
In general, non-minimality in the dark matter sector opens up new discovery prospects at the LHC. In this work, we identified a distinctive collider signature that generically emerges when the dark sector contains a $U(1)$ gauge boson. We have argued on general effective field theory grounds that such models will often give rise to final states with a dilepton resonance from the $Z'$ in association with missing energy. The example simplified models we have discussed illustrate the various mechanisms to produce the $Z'$ (in decays or radiation off of other dark sector particles), the possibilities for its decay (100\% to the Standard Model, or mostly to the dark sector), and the different kinematic regimes for the signal (such as highly boosted dark sector production through a heavy mediator, or low boost from a compressed decay). In much of model space the $Z'$ is completely hidden under SM backgrounds in standard resonance searches, but can be discovered through the proposed resonance + MET analysis.      

A search for dilepton resonances plus missing energy would be a straightforward extension of existing LHC analyses in the dilepton + MET channel. (Indeed, published data~\cite{Aad:2014vma} can already place bounds, as in Figure~\ref{fig:ATLASconstraints}.) We have seen that dark sector models can realize a range of resonance masses and boosts. An inclusive analysis which could achieve near-optimal discovery reach for this broad range of models would involve two simple extensions to existing LHC dilepton + MET search selections:

\begin{itemize}
\item{Scanning over as wide a range of resonance masses as possible, with the dilepton mass window of the signal selection chosen to optimize sensitivity to a narrow resonance}
\item{Implementing multiple signal regions corresponding to different cuts on MET and dilepton $p_T$, in particular a signal region with a hard enough cut to almost completely eliminate the SM background (which would still retain sensitivity to some signals)} 
\end{itemize}

Such a strategy should saturate the LHC's potential to discover the $Z'$ plus MET final state for essentially \emph{any} dark sector model. Indeed, in forthcoming work other authors~\cite{Autran:2015mfa, Bai:2015nfa} also propose a search of this type, motivated by other models.

Further explorations of non-minimal dark sectors could lead to other interesting signatures. For example, one may consider decays of dark scalars (directly) to the Standard Model through mixing with the Higgs. By direct analogy to the discussion we have presented, there may exist dark sector processes that could produce these scalars in association with other, invisible particles. A search for $b\bar b$ or $\gamma \gamma$ resonances in association with MET may be able to reveal such processes, even if the mixing with the Higgs is too small for the new scalar to be observable in direct $2 \rightarrow 1$ production. Note that the Higgs portal and kinetic mixing operator along with the neutrino portal $H L N$ exhaust the possibilities for renormalizable coupling of a dark sector to the SM.

If new physics does not couple to the Standard Model gauge bosons, there are a limited number of possibilities for probing it at colliders. While identifying resonance provides a very powerful tool to pick out signals from SM background, current resonance searches at the LHC do not utilize all possible, motivated signatures of possible new physics. Searches in final states such as resonances plus missing energy can help realize the full discovery potential of the LHC.

\section*{Acknowledgments}
We thank David~Curtin, Cyrus~Faroughy, Heribertus~B.~Hartanto, David~E.~Kaplan, and Tongyan~Lin for discussions. The authors were supported by the NSF under grant PHY-1214000. PS was also supported in part by NSF grant PHY-1315155 and by the Maryland Center for Fundamental Physics.

\appendix

\section{Monte Carlo Simulation}
\label{sec:MC}

To derive projected limits on new physics models, we performed Monte Carlo simulation of both signal events and Standard Model backgrounds. We generated event samples at the partonic level using MadGraph~v5~\cite{Alwall:2014hca}. For the background samples, we used the Standard Model UFO model provided with MadGraph. New physics UFO models were created using Feynrules v2.0 \cite{Alloul:2013bka}. Parton showering and hadronization were performed using Pythia v6 \cite{Sjostrand:2006za}. We used Delphes v3~\cite{deFavereau:2013fsa} to simulate detector effects including the addition of pile-up
(consisting of minimum bias events generated in Pythia), and to reconstruct jets (using FastJet~\cite{Cacciari:2011ma}), leptons (including isolation cuts), and missing transverse momentum. To subtract the neutral pile-up components we used the jet area method \cite{Cacciari:2007fd, Cacciari:2008gn} as implemented in Delphes (with the ``active'' area algorithm, and using $k_T$ jets with radius $0.6$). To reduce the effect of charged pile-up particles, ATLAS and CMS employ different strategies; we implemented Delphes modules to reproduce both of these techniques. In the ATLAS selection of ref.~\cite{Aad:2014vma}, a jet is classified as originating from pile-up if none of the tracks associated with the jet come from the primary vertex within the $z$-vertex resolution. CMS in~\cite{CMS-PAS-JME-13-005} defines the parameter $\beta^{*}$ for each jet:
\begin{equation}
\beta^* = \frac{\sum_{i \supset PU} p_T^i}{\sum_{\text{all tracks}} p_T^i},
\end{equation}
where $\sum_{i \supset PU} p_T^i$ is the $p_T$ sum of the tracks associated with the jet that do not originate from the primary vertex, while $\sum_{\text{all tracks}} p_T^i$ is the $p_T$ sum of all tracks associated with the jet (from any vertex). Pileup jets are characterized by high values of $\beta^*$. 

When simulating events at 8 TeV, we ran Delphes with the ATLAS detector card with modified charged pileup subtraction as discussed above, and with an average of 21 pileup events per crossing. For the 14 TeV projection considered in section~\ref{sec:TwinHiggs}, we simulate an average of 40 pileup events per crossing. With the increased pileup we find that signal vs. background discrimination is improved by using a variant of the CMS strategy for charged pileup subtraction, identifying jets as pileup if $\beta^* > 0.8$ (rather than $0.2$ as in the analysis of~\cite{CMS-PAS-JME-13-005}). For 14 TeV simulation therefore we ran Delphes with the CMS card, modified to implement this pileup subtraction technique.

\begin{figure}[tbp]
\begin{center}
\begin{subfigure}[b]{0.45\textwidth}
	\includegraphics[width=\textwidth]{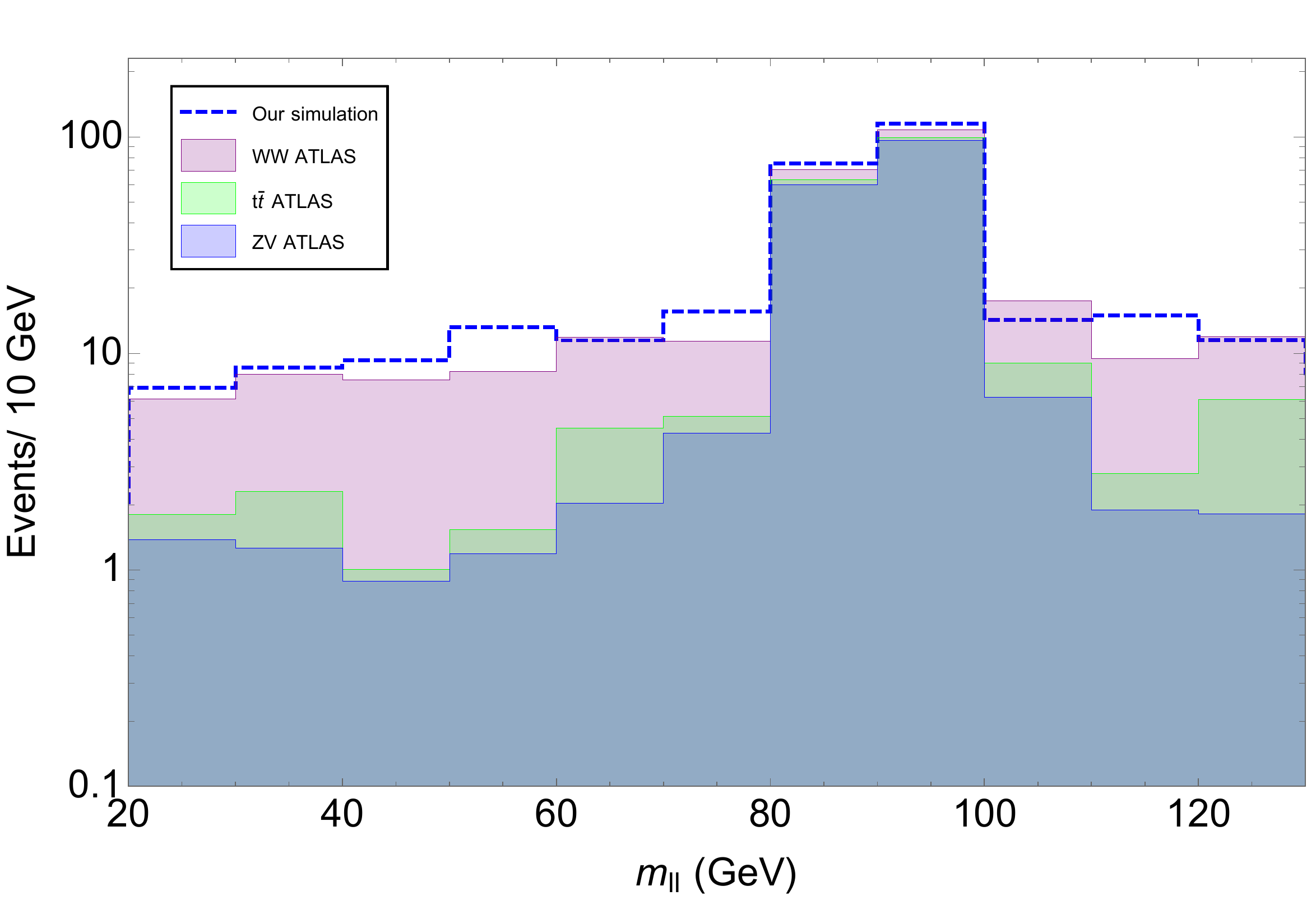}
	\caption{$m_{\ell\ell}$}
	\label{Fig:8TeVmll}
\end{subfigure}
\begin{subfigure}[b]{0.45\textwidth}
	\includegraphics[width=\textwidth]{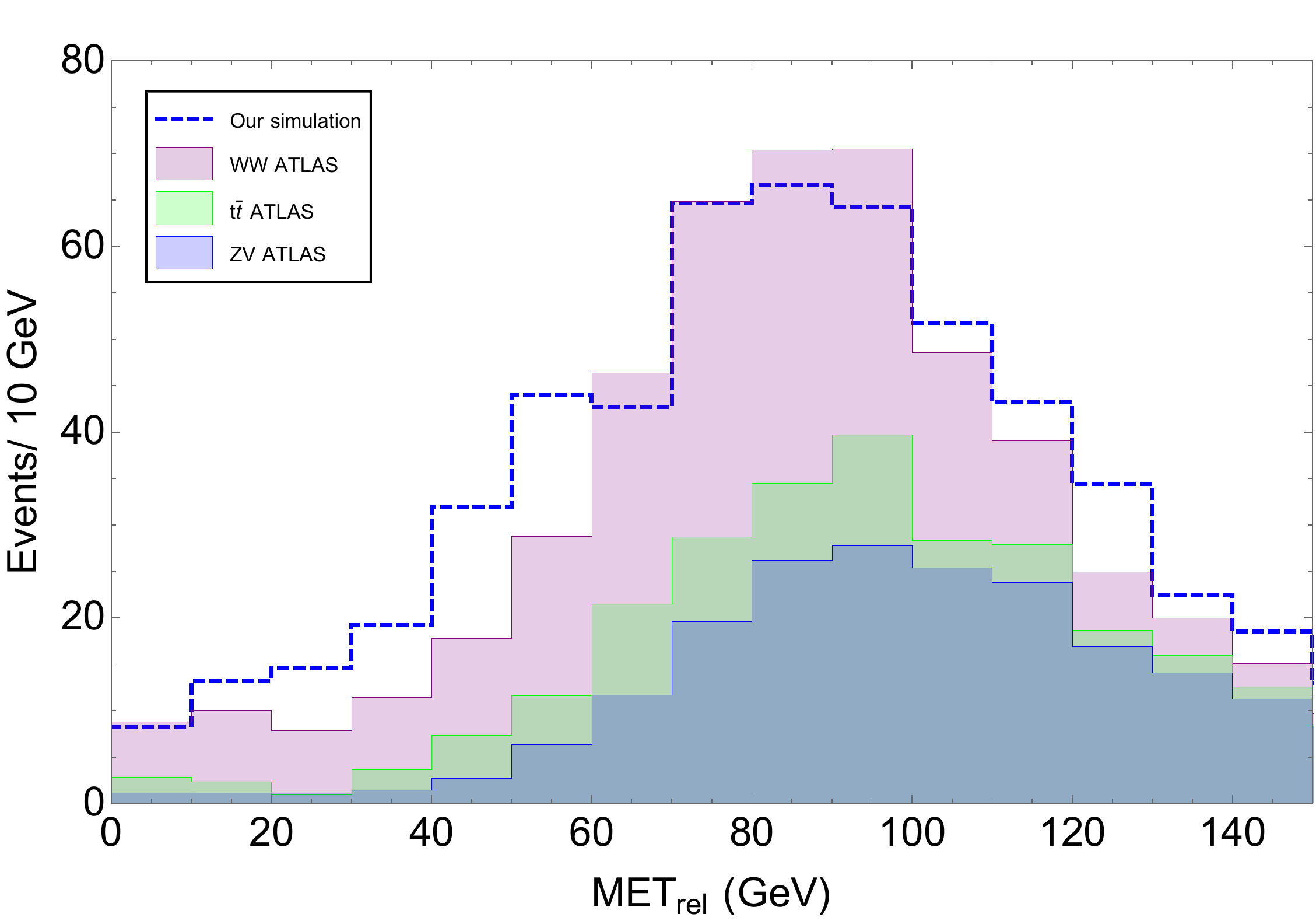}
	\caption{${\met}_{, rel}$}
	\label{Fig:8TeVmet}
\end{subfigure}
\end{center}
\caption{ Comparison between ATLAS and our simulated background for LHC 8 TeV with luminosity of 20 fb$^{-1}$.} 
\label{Fig:ATLAS8}
\end{figure}

After subtracting pileup tracks and calorimeter energy, we implemented jet finding (for jet vetos) and lepton isolation in Delphes to match the object reconstruction in the ATLAS search~\cite{Aad:2014vma}: jets are reconstructed using the anti-$k_T$ algorithm with $\Delta R = 0.4$, muons are required to have the scalar sum of $p_T$ of tracks above 400 MeV within $\Delta R < 0.3$ to be less than 16\% of the muon $p_T$, and electrons have the same isolation cut plus the requirement that the sum of $E_T$ in calorimeter clusters within $\Delta R < .3$ is less than 18\% of the electron $p_T$. 

The main SM backgrounds for our analyses are $W^+W^-$ and $t\bar t$.  $ZZ$ and $ZW$ processes also contribute (and dominate) for $m_{\ell\ell}$ near the $Z$ mass. In the signal regions of interest to us, the background contribution from processes with fake leptons and/or MET is small, as demonstrated in~\cite{Aad:2014vma}. We computed background event rates by applying selection efficiencies from our MC simulation to the total cross-sections for each process, obtained at NLO for diboson processes from~\cite{Campbell:2011bn} and at NNLL for $t\bar{t}$ from~\cite{Czakon:2013goa} (without including any dependence of the $K$-factors on the experimental cuts).  Figure~\ref{Fig:ATLAS8} shows a comparison between the results of our MC and the ATLAS simulation for certain background distributions from~\cite{Aad:2014vma}.

To obtain bounds on signal models from the published ATLAS results (as in figure~\ref{fig:ATLASconstraints}), we derive $95\%$ CL bounds from a Bayesian likelihood analysis on the binned data of Figure 3a of~\cite{Aad:2014vma}, with signal distributions as predicted by our MC and a flat prior for the signal cross-section times branching ratio. To obtain the projected bounds from the more optimized searches we propose, we apply the cuts of Table~\ref{tab:cuts} to simulated backgrounds and signal, with modified ${\met}_{rel}$, $p_{T,\ell\ell'}$ and $m_{\ell\ell'}$ cuts as discussed in section~\ref{sec:search}. From the resulting values for the expected signal efficiencies and background count rates we can derive the expectation value of the $95\%$ C.L. bound on the signal rate one would derive assuming Standard Model backgrounds only (as in figures~\ref{fig:Darkstrahlungconstraints},~\ref{fig:Cascadeconstraints},~\ref{fig:H2Zpconstraints}), or the expected significance level of the excess in this channel if a particular signal model is realized (as in figure~\ref{fig:TwinHiggsConstraints}).


\bibliography{./lit1}

\end{document}